\documentclass[aps,prb,amsmath,amssymb,twocolumn,superscriptaddress,floatfix]{revtex4-2}
\usepackage{graphicx}
\usepackage{hyperref}
\usepackage{bm}
\usepackage[dvipsnames]{xcolor}
\hypersetup{pdftitle={Rouaite},pdfauthor={Darren},pdfsubject={Frustrated magnetism},pdfdisplaydoctitle}

\def\rouaH{Cu$_\text{2}$(OH)$_\text{3}$NO$_\text{3}$}
\def\rouaD{Cu$_\text{2}$(OD)$_\text{3}$NO$_\text{3}$}
\def\botHBr{Cu$_\text{2}$(OH)$_\text{3}$Br}
\def\botDBr{Cu$_\text{2}$(OD)$_\text{3}$Br}
\def\botHCl{Cu$_\text{2}$(OH)$_\text{3}$Cl}

\def\cue{$\mathbf{q}$}

\def\TN{\ensuremath{T_\text{N}}}

\begin{document}

\title{Magnetic Phase Diagram of Rouaite, \rouaH}

\author{Aswathi Mannathanath Chakkingal}
\author{Anton A.\ Kulbakov}
\author{Justus Grumbach}
\author{Nikolai S.\ Pavlovskii}
\author{Ulrike Stockert}
\author{Kaushick K.\ Parui}
\affiliation{Institut f\"ur Festk\"orper- und Materialphysik, Technische Universit\"at Dresden, 01062 Dresden, Germany}

\author{Maxim Avdeev}
\affiliation{Australian Nuclear Science and Technology Organisation, Lucas Heights, NSW 2234, Australia}
\affiliation{School of Chemistry, The University of Sydney, Sydney, NSW 2006, Australia}

\author{R.\ Kumar}
\author{Issei Niwata}
\affiliation{Department of Physics, Faculty of Science, Hokkaido University, Sapporo 060-0810, Japan}

\author{Ellen H{\"a}u{\ss}ler}
\affiliation{Anorganische Chemie II, Technische Universit{\"a}t Dresden, 01062 Dresden, Germany}

\author{Roman Gumeniuk}
\affiliation{Institut für Experimentelle Physik, TU Bergakademie Freiberg, 09596 Freiberg, Germany}

\author{J.\ Ross Stewart}
\author{James P.\ Tellam}
\affiliation{ISIS Neutron and Muon Source, Rutherford Appleton Laboratory, Didcot OX11~0QX, UK}

\author{Vladimir Pomjakushin}
\affiliation{Laboratory for Neutron Scattering and Imaging, Paul Scherrer Institute, CH-5232 Villigen, Switzerland}

\author{Sergey Granovsky}
\affiliation{Institut f\"ur Festk\"orper- und Materialphysik, Technische Universit\"at Dresden, 01062 Dresden, Germany}

\author{Mathias Doerr}
\author{Elena Hassinger}
\affiliation{Institut f\"ur Festk\"orper- und Materialphysik, Technische Universit\"at Dresden, 01062 Dresden, Germany}

\author{Sergei Zherlitsyn}
\affiliation{Hochfeld-Magnetlabor Dresden (HLD), Helmholtz-Zentrum Dresden-Rossendorf (HZDR), 01328 Dresden,
Germany}

\author{Yoshihiko Ihara}
\affiliation{Department of Physics, Faculty of Science, Hokkaido University, Sapporo 060-0810, Japan}

\author{Dmytro S.\ Inosov}
\email{dmytro.inosov@tu-dresden.de}
\affiliation{Institut f\"ur Festk\"orper- und Materialphysik, Technische Universit\"at Dresden, 01062 Dresden, Germany}
\affiliation{W\"urzburg-Dresden Cluster of Excellence on Complexity and Topology in Quantum Matter\,---\,ct.qmat, Technische Universit\"at Dresden, 01062 Dresden, Germany}

\author{Darren C.\ Peets}
\email{darren.peets@tu-dresden.de}
\affiliation{Institut f\"ur Festk\"orper- und Materialphysik, Technische Universit\"at Dresden, 01062 Dresden, Germany}

\begin{abstract}

Spinon-magnon mixing was recently reported in botallackite Cu$_2$(OH)$_3$Br with a uniaxially compressed triangular lattice of Cu$^{2+}$ quantum spins\,\cite{Zhang2020}.  Its nitrate counterpart rouaite, Cu$_2$(OH)$_3$NO$_3$, has a highly analogous structure and might be expected to exhibit similar physics.  To lay a foundation for research on this material, we clarify rouaite's magnetic phase diagram and identify both low-field phases.  The low-temperature magnetic state consists of alternating ferro- and antiferromagnetic chains, as in botallackite, but with additional canting, leading to net moments on all chains which rotate from one chain to another to form a 90$^\circ$ cycloidal pattern.  The higher-temperature phase is a helical modulation of this order, wherein the spins rotate from one Cu plane to the next.  This extends to zero temperature for fields perpendicular to the chains, leading to a set of low-temperature field-induced phase transitions.  Rouaite may offer another platform for spinon-magnon mixing, while our results suggest a delicate balance of interactions and high tunability of the magnetism. 

\end{abstract}
\maketitle 

\section{Introduction}
In magnetically frustrated systems, the strong exchange interactions compete with each other, and the magnetic ground state can depend crucially on weaker interactions that can ordinarily be neglected, or on details of the frustration.  Such systems can be exquisitely tunable, since any minor perturbation can upset this delicate balance and send the system into a completely different magnetically ordered state.  This situation is most commonly realized by arranging the magnetic ions in a geometry that pits different interactions against each other\,\cite{Ramirez1994,Lacroix2011,Diep2013,Batista2016,Schmidt2017}, with the classic example being spins on a triangle with antiferromagnetic interactions.  The competition among interactions that destabilizes conventional forms of magnetic order can be aided by limiting the number of interactions at each site, for instance in low-dimensional systems, or by quantum fluctuations.  The latter are most relevant for small spins, especially $S=1/2$.

The stability of the $3d^9$ electronic configuration of Cu$^{2+}$ makes it a particularly accessible magnetic ion for quantum magnetism, while its low spin-orbit coupling makes it a nearly pure-spin moment.  In condensed-matter physics research, Cu$^{2+}$ is most commonly found in complex oxides, typically forming bipartite square magnetic sublattices, for instance in the cuprate superconductors\,\cite{Park1995}.  A bipartite lattice can exhibit bond frustration from a competition between nearest- and next-nearest-neighbor interactions, but it is not geometrically frustrated.  

A number of copper compounds exist in which the lattice is not bipartite, many of which are known primarily as minerals.  The copper-based minerals are often composed of distorted Cu$^{2+}$ triangles\,\cite{Inosov2018}, and have proven a rich platform for novel physics.  Examples include the candidate quantum spin-liquid state in herbertsmithite ZnCu$_3$(OH)$_6$Cl$_2$\,\cite{Shores2005,Norman2016,Lancaster2023}; enormous effective moments in atacamite Cu$_2$Cl(OH)$_3$\,\cite{Heinze2021}; and misfit multiple-\cue\ order in antlerite Cu$_3$SO$_4$(OH)$_4$\,\cite{Kulbakov2022a,Kulbakov2022b}.  Recently, spinon-magnon interactions were reported in botallackite \botHBr\,\cite{Zhang2020}, whose triangular lattice is compressed along one leg of the triangle to form spin chains.  While spinon continua and magnon branches have been observed together in other materials\,\cite{Testa2021} this is, to our knowledge, the only material in which they have been reported to coexist in the same energy range, and there is evidence that they interact with each other.  Since magnons are bosons and spinons are fermions, any mixing of these constituents of the excitation spectrum might be expected to lead to extremely exotic physics.  The possibility of such an interaction had not been considered in detail prior to Ref.~\onlinecite{Zhang2020}, but has since led to theoretical interest\,\cite{Majumdar2021,Gautreau2021}.  However, with only one potential host material known, it is difficult to examine the interactions in detail or make general statements about the effects or occurrence of spinon-magnon interactions.  

Rouaite, \rouaH, hosts a Cu$^{2+}$ sublattice highly similar to that of botallackite\,\cite{Effenberger1983}, with the most obvious difference being the replacement of Br$^-$ with NO$_3^-$.  The strongest interactions are expected to be along the crystallographic $b$ axis, leading to a proposed alternation of ferro- and antiferromagnetic chains\,\cite{Ruiz2006,Kikuchi2018}, which is thought to be the key ingredient for spinon-magnon interactions in botallackite.  Previous reports on rouaite found a N\'eel transition around 6--10\,K\,\cite{Drillon1995,Linder1995,Kikuchi2018}, a metamagnetic transition around 2\,T\,\cite{Kikuchi2018,Yuan2022}, an apparent additional transition just below 5\,K\,\cite{Yuan2022} which was only seen in the magnetic susceptibility and only for specific field orientations, and several possible transitions at high field\,\cite{Yuan2022}.  Magnetic phase diagrams have been reported on powder\,\cite{Kikuchi2018} and single crystals\,\cite{Yuan2022}.  However, neutron powder diffraction at low temperature failed to find any magnetic reflections, leading to the conclusion that any long-range order was of extremely small moments and the material was most likely a resonating valence bond system\,\cite{Yuan2022}, in which the spins would form a liquid of spin-0 dimers.  A radically different ground state from that found in structurally similar botallackite can be readily rationalized through the high tunability in frustrated low-dimensional quantum spin systems, but is at odds with the theoretical predictions\,\cite{Ruiz2006} and difficult to reconcile with previous experimental results, particularly nuclear magnetic resonance (NMR) on powder samples\,\cite{Kikuchi2018}.

Here we report a detailed magnetic phase diagram for rouaite for fields along all three principal magnetic field directions.  The magnetic ground state consists of alternating ferro- and antiferromagnetic chains as previously proposed, but with significant canting angles not reported in botallackite.  The intermediate-temperature phase, which is not observed in botallackite, is a helical modulation along $c$ of the lower-temperature phase, implying a key role for interlayer exchanges which are expected to be extremely weak.  This phase extends to zero temperature in finite fields perpendicular to $b$.  We also logically propose a plausible identity of the phase present in fields above $\sim3$\,T, which likely finds an analogy in botallackite.

\section{Experimental}

Powder samples of synthetic rouaite were prepared hydrothermally under autogenous pressure. Similar to the synthesis reported in Ref.~\onlinecite{Yoder2010b}, 5\,mL of a 1-M solution of Na$_2$CO$_3\cdot$H$_2$O (Gr\"ussing GmbH, 99.5\%) were slowly added to 10\,mL of a 1-M solution of Cu(NO$_3$)$_2\cdot 3$H$_2$O (Acros Organics, 99\%).  This was transferred to a 50\,mL Teflon-lined stainless-steel autoclave, sealed, and heated in a convection drying oven.  After heating to 180\,$^\circ$C in 2\,h, the mixture was allowed to react at that temperature for 5 days, cooled to 50\,$^\circ$C in 2\,h, then cooled freely to room temperature.  The hydrothermal reaction products consisted of light blue supernatant (pH 5--6) and turquoise rouaite powder, which was filtered, washed with deionised water, and dried at room temperature.  For the deuterated samples required to minimize incoherent neutron scattering, the precusor solutions were made using D$_2$O (Acros Organics, 99.8\,at.\,\% D). 

\begin{figure*}
  \includegraphics[width=\textwidth]{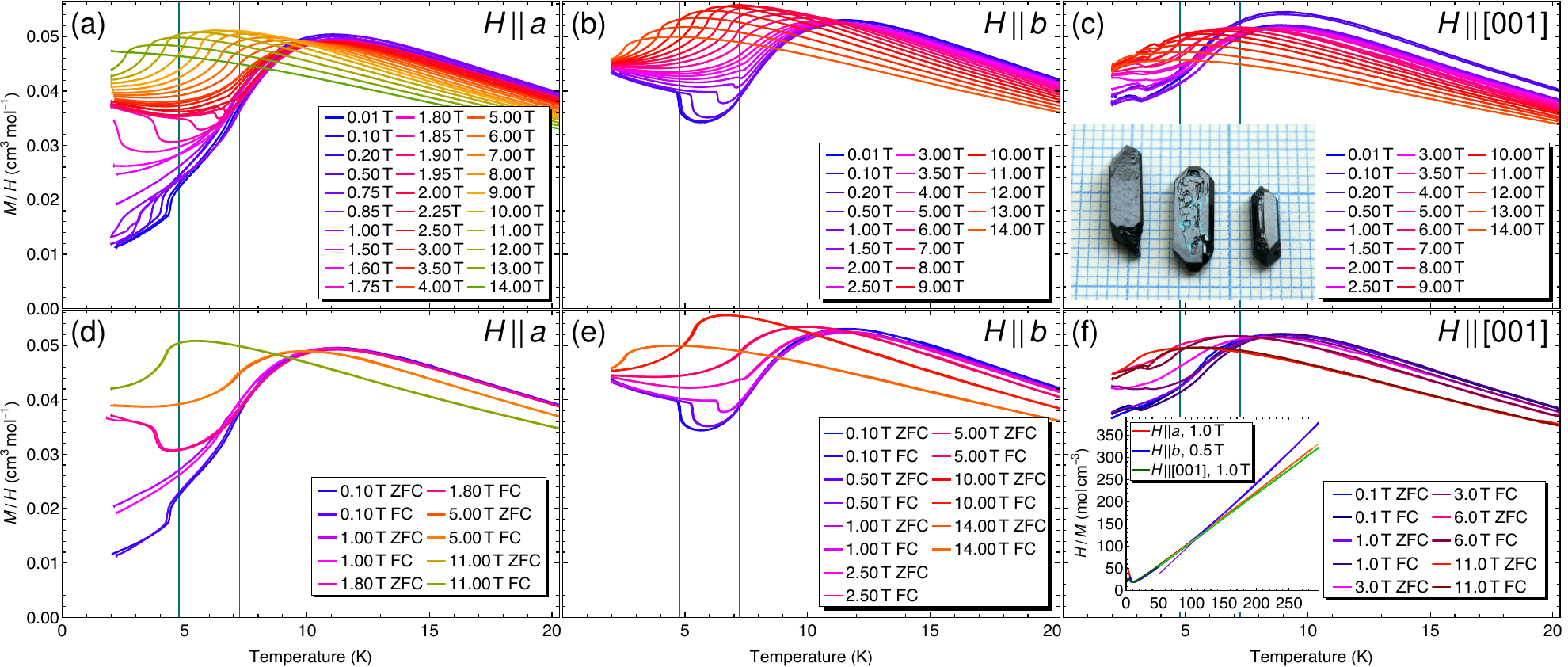}
  \caption{\label{MvsT}Selected magnetization data on rouaite.  Field-cooled data are shown for the three mutually orthogonal directions (a) $H\parallel a$, (b) $H\parallel b$ and (c) $H\parallel [0\,0\,1]$.  At selected fields, these data are compared against zero-field-cooled data for (d) $H\parallel a$, (e) $H\parallel b$ and (f) $H\parallel [0\,0\,1]$.  The transitions found at zero field in the specific heat are marked with vertical lines.  The inset to (c) shows several crystals on mm-ruled graph paper, while the inset to (f) plots the inverse magnetization to higher temperatures, together with Curie-Weiss fits.}
\end{figure*}

Single crystals were grown from a more-concentrated solution of Cu(NO$_3$)$_2\cdot 3$H$_2$O~:~H$_2$O = 8:1 by mass, and a higher-temperature PPL liner was used.  In this case, the mixture was held at 240\,$^\circ$C for 5 days.  The supernatant was a dark blue-green (pH 2--3), and contained large deep-blue rouaite crystals of size up to 12$\times$5$\times$5\,mm$^3$ [Fig.\ \ref{MvsT}(c) inset], which were washed with water and dried at room temperature.  Starting from undeuterated copper nitrate trihydrate and D$_2$O would limit the deuteration to 60--65\% in such a synthesis, and water-free Cu(NO$_3$)$_2$ is not accessible as it decomposes upon heating, so it was essential to prepare deuterated precursor [Cu(D$_2$O)$_6$](NO$_3$)$_2$.  Cu(NO$_3$)$_2\cdot 3$H$_2$O was dissolved in D$_2$O and then repeatedly recrystallized by slowly evaporating the solvent in a rotary evaporator in order to ensure a high D content in our precursor nitrate.  The synthesis then proceeded as for the undeuterated crystals.

Neutron diffraction data on deuterated rouaite powder were collected from 3.9$^\circ$ to 163.8$^\circ$ using 2.44-\AA\ neutrons at the Echidna\,\cite{Echidna} diffractometer and from 16$^\circ$ to 133.75$^\circ$ with 2.41-\AA\ neutrons at the Wombat\,\cite{Wombat} diffractometer, both at the OPAL research reactor, Australian Science and Technology Organization (ANSTO), Lucas Heights, Australia.  Further data were collected for significantly shorter count times at the HRPT diffractometer, SINQ, PSI, Switzerland\,\cite{HRPT_PSI}, at wavelengths of 1.89 and 2.45\,\AA.  Additional neutron scattering on a $\sim$1.7-g mosaic of single crystals was performed on the LET\,\cite{Bewley2011} time-of-flight spectrometer at the ISIS neutron source, Didcot, UK, using incoming neutron energies of 20.03, 5.96, 2.82, and 1.64\,meV.  Powder diffraction data were Rietveld-refined in {\sc FullProf} by the full-matrix least-squares method\,\cite{FullProf}, using neutron scattering factors from Ref.~\onlinecite{Sears1992}.

Neutron Laue diffraction patterns of a $\sim$1\,mm$^3$ rouaite single crystal were measured for ten distinct sample orientations with respect to the incident neutron beam at 10, 5.5, and 2.1\,K using the Koala2 white-beam neutron Laue diffractometer\,\cite{Koala} at the OPAL Research Reactor, Australian Centre for Neutron Scattering (ACNS), Australian Nuclear Science and Technology Organisation (ANSTO), in Sydney, Australia. Image data processing, including indexing, intensity integration, and wavelength distribution normalization, was performed using LaueG\,\cite{LaueG}.  Crystal and magnetic structure refinements were carried
out using {\sc Jana2006}\,\cite{Jana2006} and checked using {\sc FullProf}\,\cite{FullProf}.

Magnetization measurements were performed by vibrating sample magnetometry (VSM) in a Cryogenic Ltd.\ Cryogen-Free Measurement System (CFMS), under zero-field-cooled-warming, field-cooled-cooling, and field-cooled-warming conditions.  Four-quadrant $M$--$H$ loops were measured at several temperatures. The single crystals were mounted to a plastic bar using GE varnish.  The same apparatus was used to collect ac susceptometry data as a function of field and temperature, using an ac field of 0.1\,mT at several frequencies, while sweeping the sample vertically through the coils at 0.05\,Hz;  3--5 such spatial oscillations were averaged.  

Low-temperature specific heat measurements were performed on a single crystal using a Physical Property Measurement System (PPMS) DynaCool-12 from Quantum Design, equipped with a $^3$He refrigerator. Measurements were taken using both $^3$He and $^4$He specific heat pucks.  Contributions from the sample holder and Apiezon N grease were subtracted. Multiple data points were collected at each temperature and averaged; the first data point at each temperature was discarded to exclude the possibility of incomplete thermal stabilization. A few additional datasets were collected in a similar manner on a CFMS using the relaxation calorimetry option.

Dilatometry measurements were performed using a tilted-plate capacitive dilatometer with a sensitivity to relative length changes of $\sim$10$^{-7}$~\cite{Rotter1998}, which was mounted on an Oxford Instruments $^4$He-flow cryostat equipped with a superconducting magnet capable of fields up to 10\,T.  The sweep rate of the magnetic field was chosen to be between 0.05 and 0.25\,T/min. For accurate monitoring and control of the dilatometer and sample temperature we used a Cernox thermometer attached to the dilatometer cell close to the sample.  Measurements of magnetostriction and thermal expansion were made on single crystals for length changes parallel or perpendicular to the mutually orthogonal crystallographic $a$, $b$, or [0\,0\,1] directions for magnetic fields oriented along each of these directions. The longitudinal and transverse components of the striction tensor found in this way allow the distortions and volume effects of the crystal lattice to be calculated. This allows one to identify all magnetic transitions accompanied by lattice effects through dilatometry, which can hint at modifications of the magnetic structure.

Ultrasound measurements were performed in magnetic fields applied along the $b$ axis using a pulse-echo method with a phase-sensitive detection technique\,\cite{Zherlitsyn2014,Luethi2005}. The polarization and wavevector of the sound waves were also along $b$.  To generate and detect ultrasound signals in the 20--120\,MHz range, we bonded ultrasound transducers (LiNbO$_3$, 36$^\circ$ Y-cut for longitudinal acoustic modes) with Thiokol to two parallel sample surfaces. Typically, several ultrasound echoes due to multiple propagations and reflections in the sample were observed. We used a PPMS combined with an ultrasound setup.

The low-temperature thermal conductivity $\kappa$ of rouaite was measured for in-plane heat currents, in magnetic fields along $a$, $b$, and [0\,0\,1] as a function of temperature and field, by a four-point steady-state method using two Cernox thermometers and a resistive heater.

Proton nuclear magnetic resonance ($^1$H-NMR) spectra were measured at a resonant frequency of 31.51\,MHz in magnetic fields applied along the $a$ direction.  Spectra were collected by sweeping field at a fixed frequency, and the intensity was recorded during the field sweep.  


\section{Crystal Structure}

\begin{figure}[tb]
  \includegraphics[width=\columnwidth]{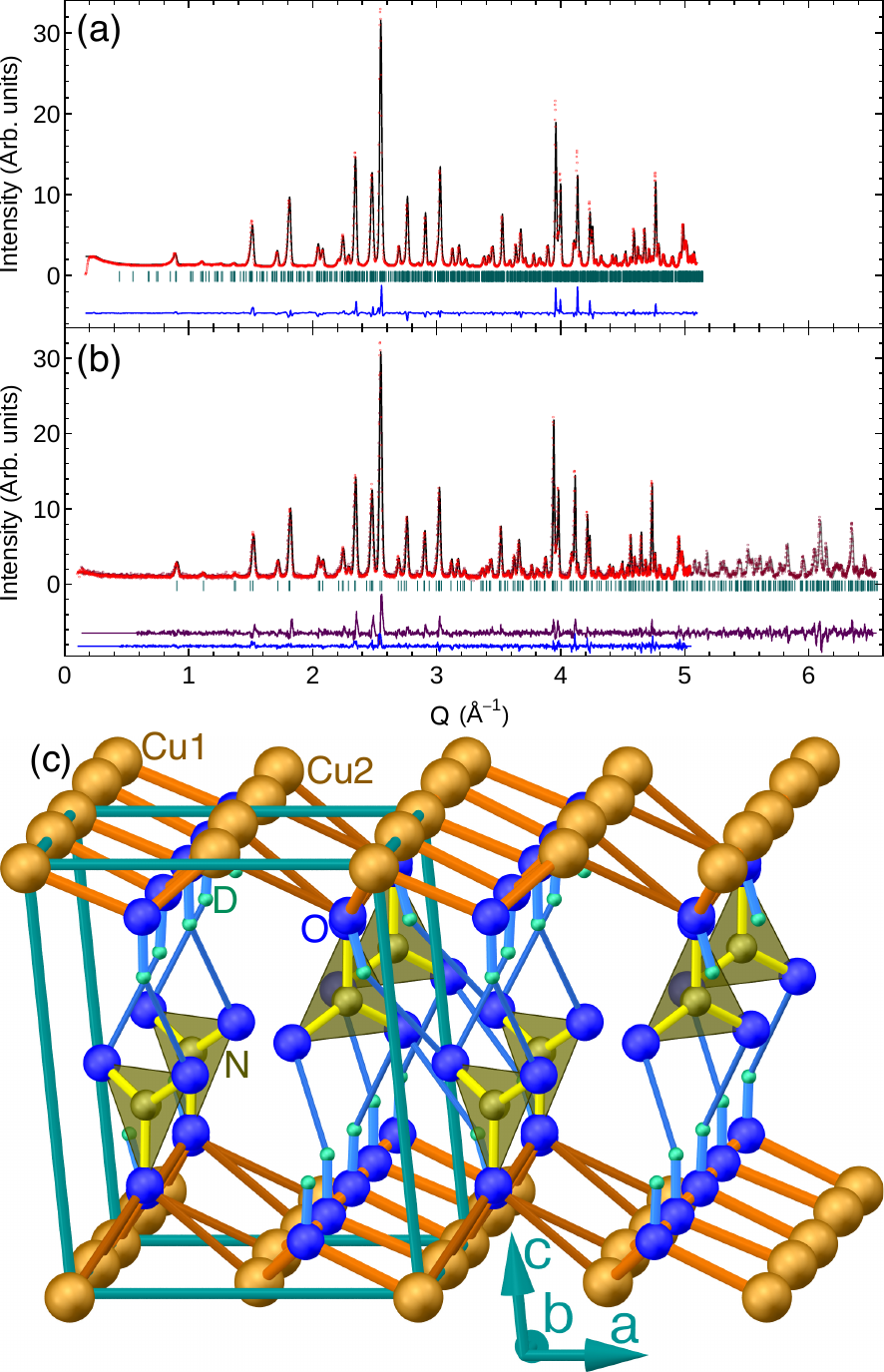}
  \caption{\label{NPD}(a) Neutron powder refinement of the crystal structure of rouaite, \rouaD\ at 20\,K based on Echidna data. (b) Joint refinement of data at 1.89 (dark) and 2.45\,\AA\ (light) at HRPT, collected at 10\,K.  (c) Refined crystal structure in monoclinic $P2_1$ from Echidna data.}
\end{figure}

Rouaite crystallizes in the monoclinic space group $P2_1$ (\#~4), with a monoclinic angle $\beta$ of roughly 95$^\circ$, with distorted-triangular-lattice Cu planes as depicted in Fig.~\ref{NPD}.  Since previous structure reports\,\cite{Nowacki1951,Nowacki1952,Effenberger1983,Pillet2006,Yuan2022} were based on x-ray diffraction, which has limited sensitivity to hydrogen positions, we start with neutron powder diffraction to verify the crystal structure and determine accurate hydrogen positions.  We previously found this to be crucial in antlerite, where calculations based on literature hydrogen positions converged on an incorrect magnetic ground state\,\cite{Kulbakov2022a}.  A Rietveld refinement was performed on a dataset collected on deuterated rouaite powder at 20\,K on the Echidna
beamline at ANSTO, Australia, and a separate joint refinement was performed on data collected at 10\,K with two neutron wavelengths at the HRPT diffractometer at PSI, Villigen, Switzerland.  The resulting refinements are shown in Figs.~\ref{NPD}(a) and \ref{NPD}(b), respectively, and the crystal structure based on the Echidna refinement is depicted in Fig.~\ref{NPD}(c).  For the Echidna refinement it was necessary to include a contribution from $\frac{\lambda}{2}$ at the 0.13\%\ level.  Details of these refinements are provided in Tables \ref{NPDsummary}, \ref{NPDEch1}, and \ref{NPDHRPT1} in Appendix~\ref{appA}, and crystallographic information files (CIFs) are available in the ancillary files online, see Appendix~\ref{supp}.  We also provide a structure refinement based on neutron Laue diffraction data, as described below.  As previously reported, rouaite is composed of distorted-triangular-lattice planes of Cu$^{2+}$ which are better viewed as coupled one-dimensional chains, while consecutive planes are linked through a network of hydrogen bonds.  This is expected to lead to a hierarchy of magnetic interactions.

\section{Magnetization}

\begin{figure*}
  \includegraphics[width=\textwidth]{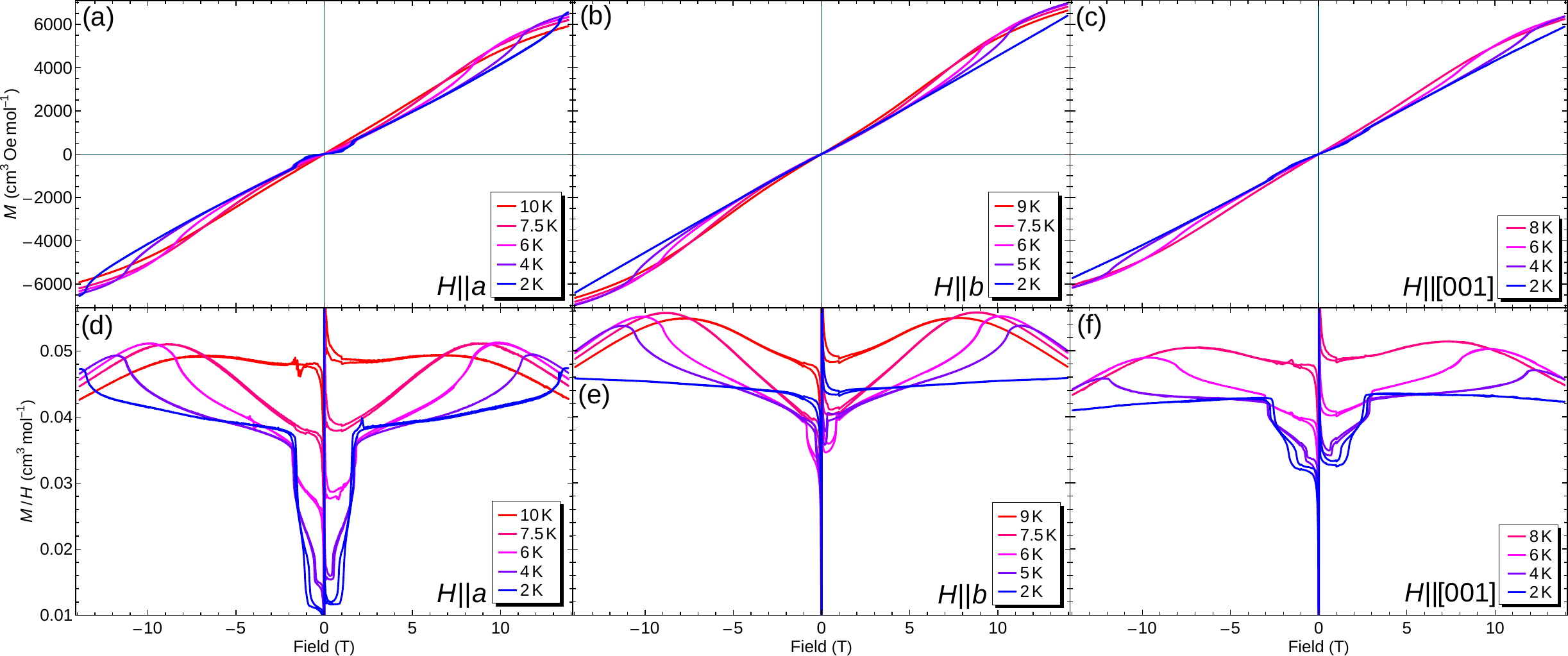}
  \caption{\label{MvsH}Selected field-dependent magnetization data on rouaite. Data are shown for (a) $H\parallel a$, (b) $H\parallel b$ and (c) $H\parallel [0\,0\,1]$, and replotted as $M/H$ in (d), (e) and (f), respectively.  Glitches at $\pm$1\,T are artefacts.}
\end{figure*}

Temperature-dependent field-cooled magnetization ($M$) data are shown for applied magnetic fields $H\parallel a$, $b$ and [0\,0\,1] in Figs.~\ref{MvsT}(a), \ref{MvsT}(b) and \ref{MvsT}(c), respectively.  For all field directions, the magnetization decreases below a peak around 10\,K, indicating that there is no net ferromagnetic moment.  The antiferromagnetic transition \TN\ would normally be defined at the inflection point below this peak, but the data for $H\parallel [0\,0\,1]$ are not consistent with the other two directions, as will be discussed later.  At low fields, below the antiferromagnetic transition the magnetization decreases abruptly around 4.7\,K for $H\parallel a$ and increases for $H\parallel b$, this temperature being consistent with the specific heat transition described below.  As field is increased, this transition moves to higher temperature for $H\parallel b$, ultimately merging with \TN\ around 2.5\,T.  For $H\parallel a$ this transition instead collapses to zero temperature around 1\,T, becoming strongly hysteretic.  The increase in $T$ with field for $H\parallel b$ is in stark contrast to published phase diagrams for the other two field orientations\,\cite{Yuan2022}.  

A second transition appears below \TN\ for $H\parallel a$, first appearing around 1.75\,T.  It nearly reaches \TN\ by 2\,T, and seemingly turns upward to merge with \TN\ around 2.5\,T, although it becomes difficult to determine \TN\ where these transitions come together.  

For $H\parallel [0\,0\,1]$, no clear transitions were seen in temperature sweeps, even at \TN.  Sweeps with $H \parallel b$ had a maximum slope around \TN\ and for $H \parallel a$ this transition appears as a weak, broadened step, but it was not possible to define any transitions for $H \parallel [0\,0\,1]$.  Sharp features were, however, found in field sweeps, as seen in Fig.~\ref{MvsH}(c,f).  As shown in Fig.~\ref{MvsT}(f), measurements taken below \TN\ on warming and cooling in this field orientation often disagreed, with significantly lower magnetization values for cooling runs, but the magnetization measured on cooling recovered to rejoin the warming data around $\sim$3--4\,K.  The temperature associated with this recovery was not found to depend systematically on field or cooling rate, and is not believed to be a phase transition.  It is likely either a metastable training of the magnetic order or an experimental artefact arising from the less-stable sample mounting required for this orientation.  Zero-field-cooled and field-cooled warming data (not shown) agreed more closely, indicating that the differences seen in Fig.~\ref{MvsT}(f) arise from the sweep direction and not from field training.

Zero-field-cooled and field-cooled magnetization data for fields along $a$ and $b$ are compared in Figs.~\ref{MvsT}(d) and \ref{MvsT}(e), respectively.  The transitions below \TN\ for $H\parallel a$ are hysteretic, but no other significant differences were observed with field training.  These transitions are probably first order.  

The magnetization has a broad peak above \TN\ around 11--12\,K in low $ab$-plane fields, consistent with most previous reports on \rouaH\ powder\,\cite{Drillon1995} and single-crystalline samples\,\cite{Yuan2022}.  The peak is significantly lower for fields perpendicular to this plane.  This is well above \TN, and the downturn on cooling below the peak is presumably associated with short-range order.  One earlier report on single crystals placed this peak around 8\,K\,\cite{Linder1995} for all three field orientations, and the origin of this discrepancy is not understood. Reference~\onlinecite{Linder1995} did not observe sharp features below \TN\ which, combined with the suppressed and isotropic onset of order, may indicate sample quality, crystallinity, twinning, or alignment issues.  The synthesis reported in Ref.~\onlinecite{Linder1995} used a 2:1 ratio of Cu(NO$_3$)$_2$ to Mg(NO$_3$)$_2$, so the broadening and suppression of the magnetic features observed in these single crystals may indicate disorder due to incorporation of trace nonmagnetic Mg$^{2+}$ on the Cu$^{2+}$ site despite this not being detected. It is known that the complete (Mg,Cu)$_\text{2}$(OH)$_\text{3}$NO$_\text{3}$ substitution series exists, with the cation stoichiometry controlled by the Cu$^{2+}$:Mg$^{2+}$ ratio during synthesis, and that Mg$^{2+}$ doping rapidly suppresses the antiferromagnetic downturn in the magnetization\,\cite{Atanasov1994}.  Mg$^{2+}$ preferentially occupies the Cu2 site\,\cite{Atanasov1994}, which is proposed to form antiferromagnetic chains\,\cite{Ruiz2006}. 


Measurements to higher temperature are plotted as inverse magnetization in the inset to Fig.~\ref{MvsT}(f), along with Curie-Weiss fits to the data above 100\,K.  The inverse magnetization for $b$-axis fields is slightly nonlinear, making extracted values less reliable for this field orientation, but we did not find a better fit function that would produce physical insight.  Extracted Curie-Weiss temperatures are 1.0, 17 and 0.28\,K for fields along $a$, $b$, and [0\,0\,1], respectively, all of which indicate net ferromagnetic interactions.  For fields along $a$ and [0\,0\,1] the small Curie-Weiss temperatures compared to \TN\ indicate the presence of additional antiferromagnetic interactions, and the magnetization more closely resembles that of an antiferromagnet than a ferromagnet.  Our Curie-Weiss temperatures are higher than those reported in Ref.~\onlinecite{Yuan2022}, where they were slightly negative for $H\parallel a$ and [0\,0\,1].  This is likely due to the narrower temperature range of our fit combined with slight curvature in the data, and extending our fit to lower temperatures indeed reduces the extracted Curie-Weiss temperatures.

Field-dependent magnetization data for $H\parallel b$, shown in Fig.~\ref{MvsH}(b) and as $M/H$ in Fig.~\ref{MvsH}(e), show no transition up to 14\,T at low temperature, but they do show a low-field transition between 4.7\,K and \TN. For the other two field orientations, shown in Figs.~\ref{MvsH}(a,c) and replotted as $M/H$ in Figs.~\ref{MvsH}(d,f), two separate hysteretic transitions are found at low temperatures below 3\,T.  For $H\parallel a$ these are consistent with the transitions found in the temperature-dependent magnetization.  As mentioned above, no transitions were visible for $H\parallel [0\,0\,1]$ in temperature sweeps, but these transitions are evidently present.  Aside from hysteresis at the transitions, our field-dependent data show no evidence for field training.  

\begin{figure}[t]
  \includegraphics[width=\columnwidth]{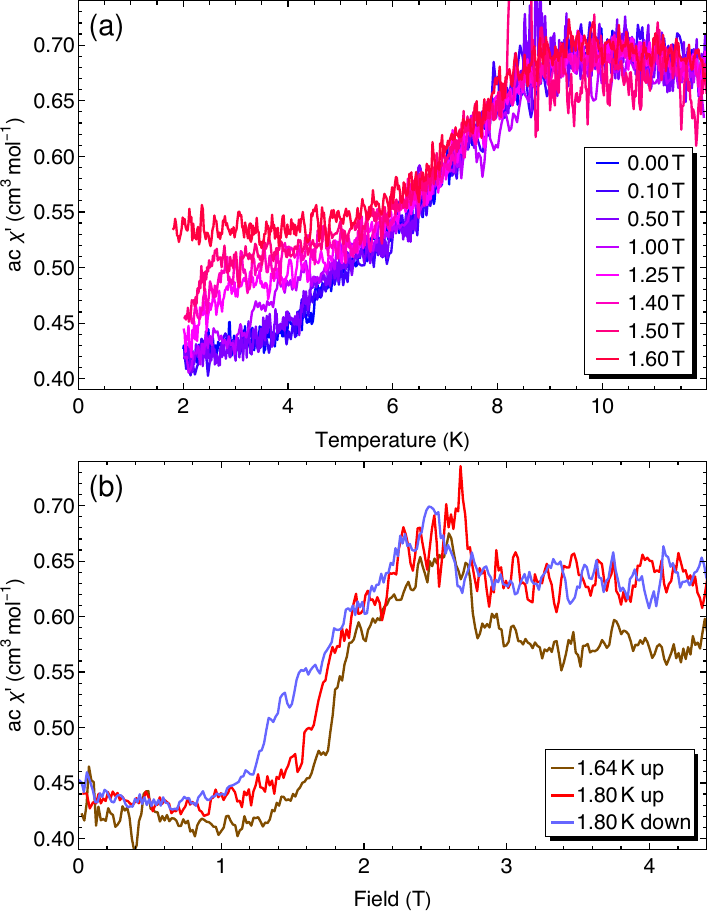}
  \caption{\label{chi}Real component of the ac susceptibility of rouaite at several (a) fields and (b) temperatures, for $H\parallel [0\,0\,1]$.  A 3-point moving average has been applied to reduce noise.}
\end{figure}

Because the magnetic transitions for $H\parallel [0\,0\,1]$ were visible in field-dependent but not temperature-dependent magnetization measurements, we also measured the ac susceptometry $\chi$ for this field direction.  Temperature sweeps of the real component $\chi^\prime$ at low fields, shown in Fig.~\ref{chi}(a), exhibit a weak step at low temperatures consistent with the transition found in field-sweep magnetization measurements, but slightly lower in temperature than in the specific heat.  The maximum slope extracted from these data returns a more plausible \TN\ than the magnetization data, but with significant uncertainty.  We did not find evidence for frequency dependence that would indicate a glass transition.  Strong steps are also observed in field sweeps, as shown in Fig.~\ref{chi}(b).  The imaginary component of the susceptibility $\chi^{\prime\prime}$ did not exhibit features above the noise level.  

\begin{figure}[b]
  \includegraphics[width=\columnwidth]{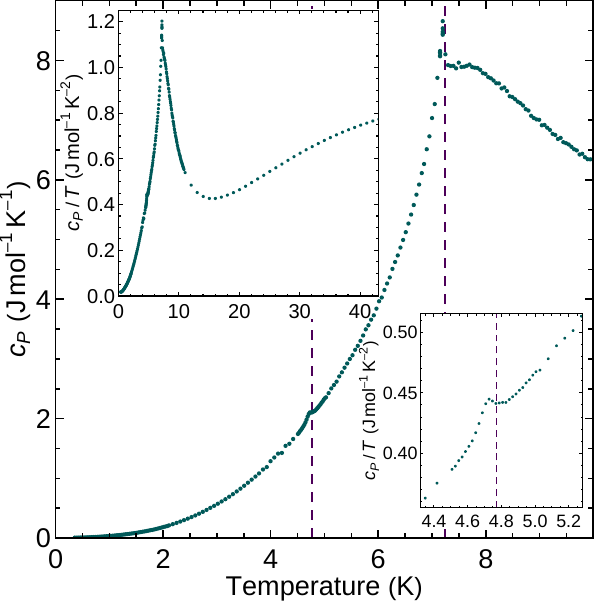}
  \caption{\label{cP}Specific heat of rouaite at zero field.  Insets show $c_P/T$ to higher temperature, demonstrating the persistence of magnetic entropy well above the 7.24-K transition, and an expanded view of the transition at 4.72\,K.  Dashed lines mark the transition temperatures.}
\end{figure}

\section{Specific Heat}

The main N\'eel transition \TN\ is found at 7.24\,K in the specific heat (Fig.~\ref{cP}), and a broad hump above that transition is presumably associated with short-range order.  The upper inset in Fig.~\ref{cP} presents $c_P/T$, where an upward deviation from the expected phonon behavior, attributed to magnetic entropy, can be discerned below $\sim$20\,K, roughly triple the ordering temperature.  This appearance of magnetic entropy well above the long-range-order transition is likely a consequence of both frustration and low dimensionality.

The lower-temperature transition was not visible in previous specific heat ($c_P$) measurements\,\cite{Wu2011,Yuan2022}.  As can be seen in Fig.~\ref{cP} this is because this transition, despite being most likely first order, is very weak and is not associated with a change in slope or power law.  The lower inset in this figure shows an expanded view of this weak peak, which is located at 4.72\,K.  Integrating this peak leads to an estimate of 2.60\,mJ\,mol$^{-1}$\,K$^{-1}$ per formula unit, or 1.30\,mJ\,mol$^{-1}$\,K$^{-1}$ per Cu, for its entropy.  If this transition is indeed first order as implied by the hysteretic magnetization data, its latent heat is very small and the phases cannot differ significantly in entropy.  This weak peak has an even weaker hump on its high-temperature side, which suggests that it could actually be a second transition at roughly 4.80\,K.  Such a small separation would likely not have been resolved in the magnetization data, particularly given that the transition is hysteretic.  

\begin{figure}[t]
  \includegraphics[width=\columnwidth]{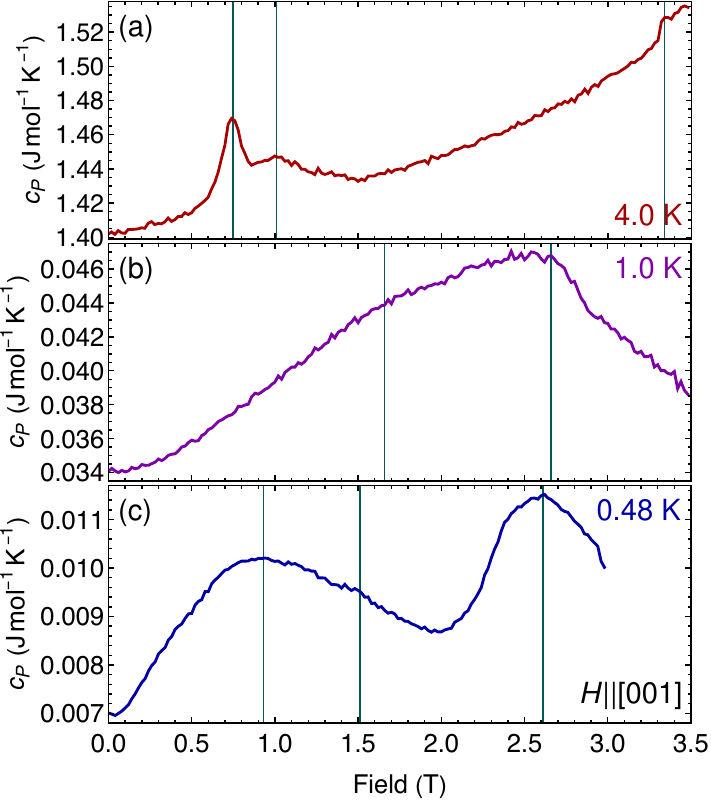}
  \caption{\label{cvsH}Field-dependent specific heat at several temperatures as labelled. Data in panel (b) were collected on decreasing field, the others on increasing field.}
\end{figure}

Additional specific heat data were collected as a function of field for $H\parallel [0\,0\,1]$, as shown for several temperatures in Fig.~\ref{cvsH}.  Here, most prominently at 4\,K, we see additional evidence suggestive of a splitting of the lower transition.  However, since no similar splitting of the transition was observed by any other technique, we cannot be certain that this splitting is real.  As mentioned, the main issue is the hysteresis observed in most other techniques.  By their nature, relaxation-time specific heat measurements oscillate the temperature, allowing the system to settle and reducing the effect of hysteresis after the first data point, which we discarded.  

An extrapolation of $cP/T$ vs $T^2$ to $T=0$\,K gives an intercept of roughly 0.012\,Jmol$^{-1}$K$^{-2}$. We do not view this as clear evidence for an additional contribution from low-energy spinons.  The spinon continuum reported in botallackite did not extend down to zero energy, having a bottom instead around 4\,meV.  We would not expect a similar spinon continuum in rouaite to be populated in the low-temperature limit.

\section{Negative Thermal Expansion}

\begin{figure}[tb]
    \includegraphics[width=\columnwidth]{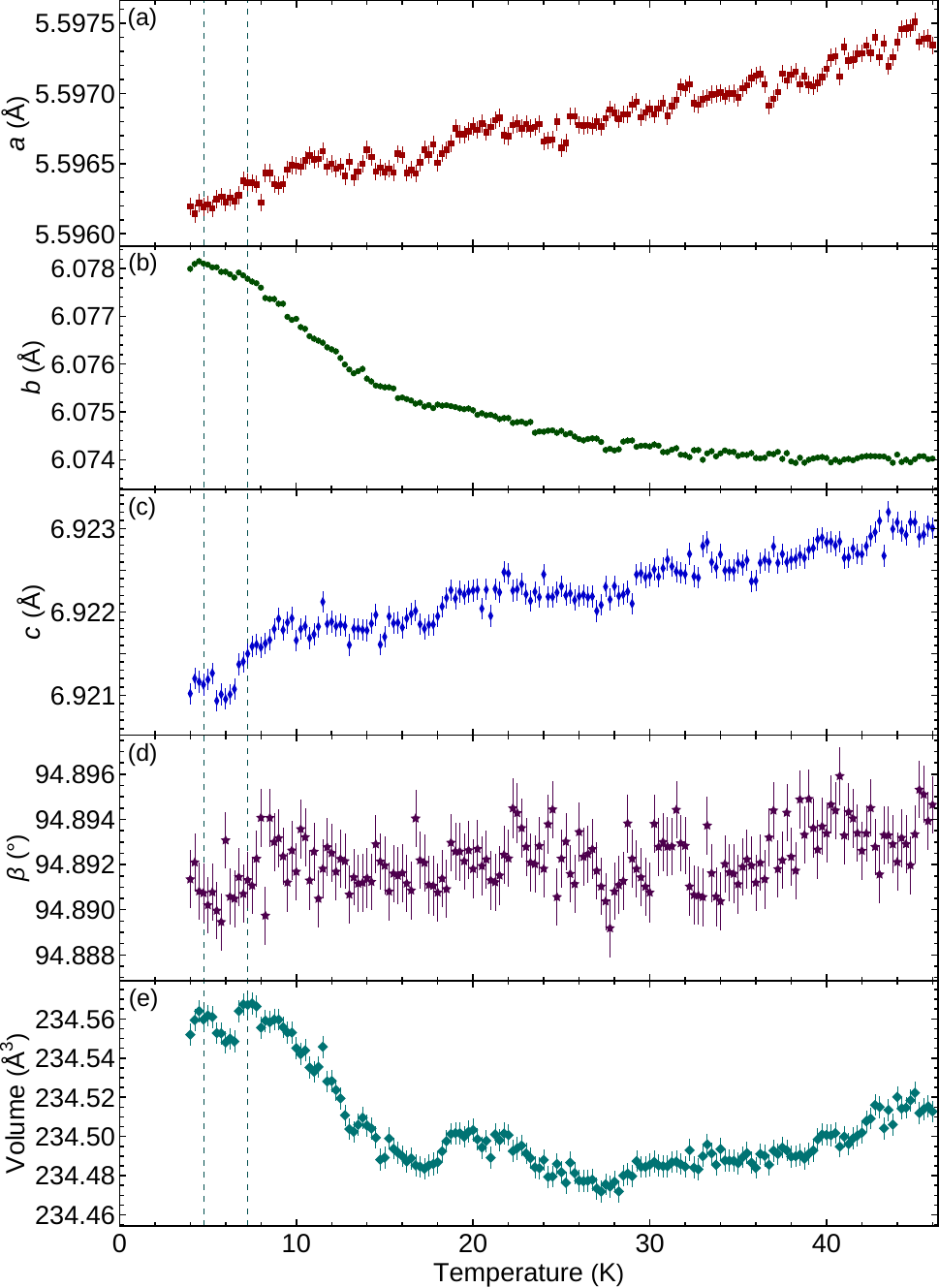}
    \caption{\label{WombatExp}Refined cell parameters and volume based on Wombat data collected at a series of temperatures.}
\end{figure}

While collecting neutron diffraction data to identify the magnetic ground state, we encountered significant shifts in a few structural Bragg peaks at low temperature, which complicated the subtraction of paramagnetic-state data to isolate the magnetic reflections.  This was unexpected --- thermal expansion is ordinarily relatively small at low temperatures, especially for a difference of less than 20\,K among the datasets.  Even more surprisingly, the peak shifts suggested negative thermal expansion.  To verify and characterize this, we collected additional data in steps of 0.25\,K from 4 to 46\,K on the high-intensity Wombat diffractometer at ANSTO, using 2.417-\AA\ neutrons.  Individual refinements at each temperature allowed us to track the temperature evolution of the lattice constants, atomic positions, and other refined parameters.  No significant, systematic trends were seen in the atomic positions within the unit cell, so these were fixed to their previously refined values.  In the final step, only the lattice parameters, background, temperature factors for Cu and O, an overall scale, and 2 peak shape parameters were refined for each temperature.  The temperature dependence of the lattice parameters based on these refinements is shown in Fig.~\ref{WombatExp}.  

Our neutron diffraction data (Wombat) evidence a substantial expansion of the $b$ axis on cooling below $\sim$40\,K, which is not compensated by contraction along the other directions. This leads to negative volume expansion below $\sim$25\,K, more than triple \TN.  This negative volume expansion saturates around the first magnetic transition at 7.24\,K.  Aside from exchange striction, crystal field striction on the Jahn-Teller Cu$^{2+}$ sites would also be possible in this system, but the long direction of the Cu2 octahedra lies along (1\,0\,$\overline{1}$) and the long axes of the Cu1 octahedra point along cell diagonals, so it is unlikely that this would lead to expansion predominantly along $b$.  Given that clear hints of magnetic entropy appear in the specific heat below $\sim$20\,K, similar to the onset in the $b$-axis thermal expansion, and given the latter's saturation near the magnetic ordering temperature, the negative thermal expansion is presumably correlated with the formation of short-range magnetic order.  The $b$ axis is the Cu chain direction, which is expected to have the strongest magnetic interactions, so its negative thermal expansion is presumably driven by magnetic interactions along the chains.  Lengthening the chains on cooling would drive the Cu--O--Cu bond angle closer to 180$^\circ$, stabilizing antiferromagnetic interactions.  

It is also worth noting here that the strong magnetoelastic coupling, implying magnetic interactions which are significant compared to the Coulomb energy scales holding the lattice together, means that the material would be expected to be highly tunable through uniaxial or hydrostatic pressure.  It is also likely crucial that calculations of the magnetic interactions begin from the low-temperature structure, and that relaxation calculations of the structure include magnetic interactions.

\begin{figure}[tb]
    \includegraphics[width=\columnwidth]{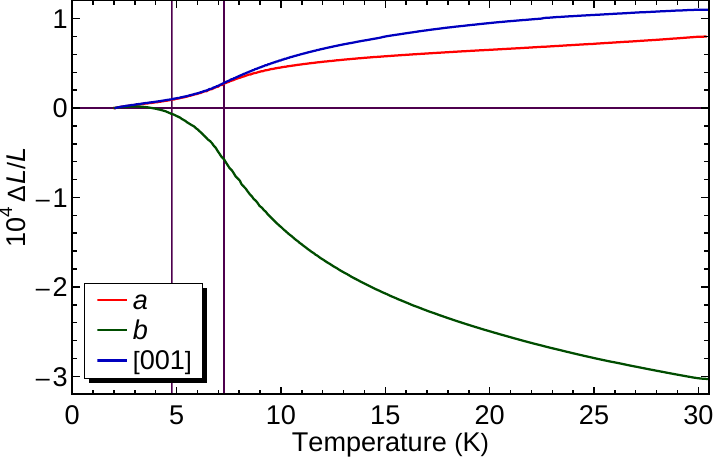}
    \caption{\label{dil_axes}Relative change of rouaite's lattice parameters with temperature, from dilatometry.}
\end{figure}

\begin{figure}[tb]
    \includegraphics[width=\columnwidth]{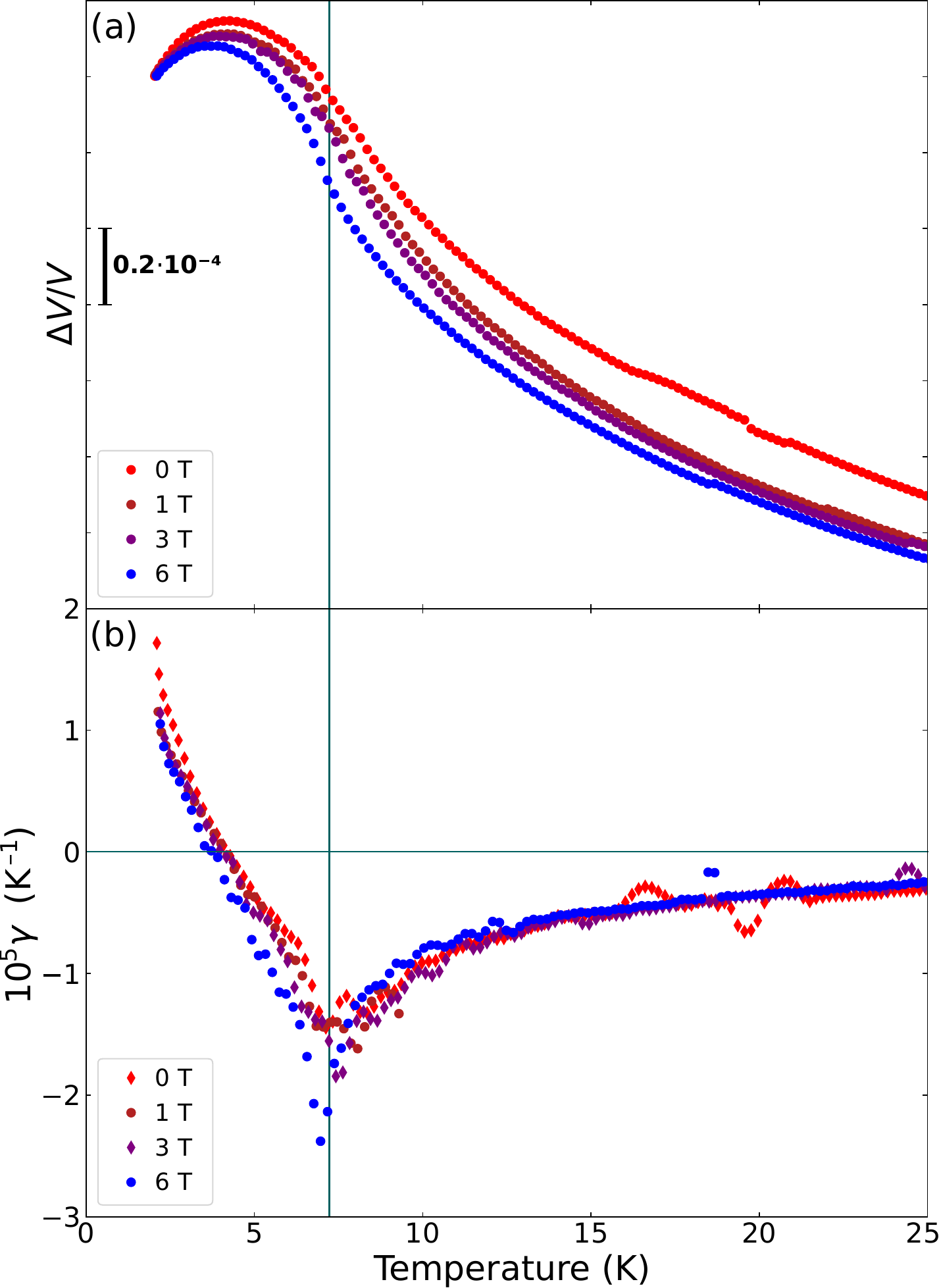}
    \caption{\label{dil_vol} (a) Field dependence of the volume expansion from dilatometry in fields $H\parallel b$. (b) Volumetric thermal expansion coefficient $\gamma$ based on the data in (a).  \TN\ at zero field extracted from $c_P(T)$ is marked.}
\end{figure}

\begin{figure*}
  \includegraphics[width=\textwidth]{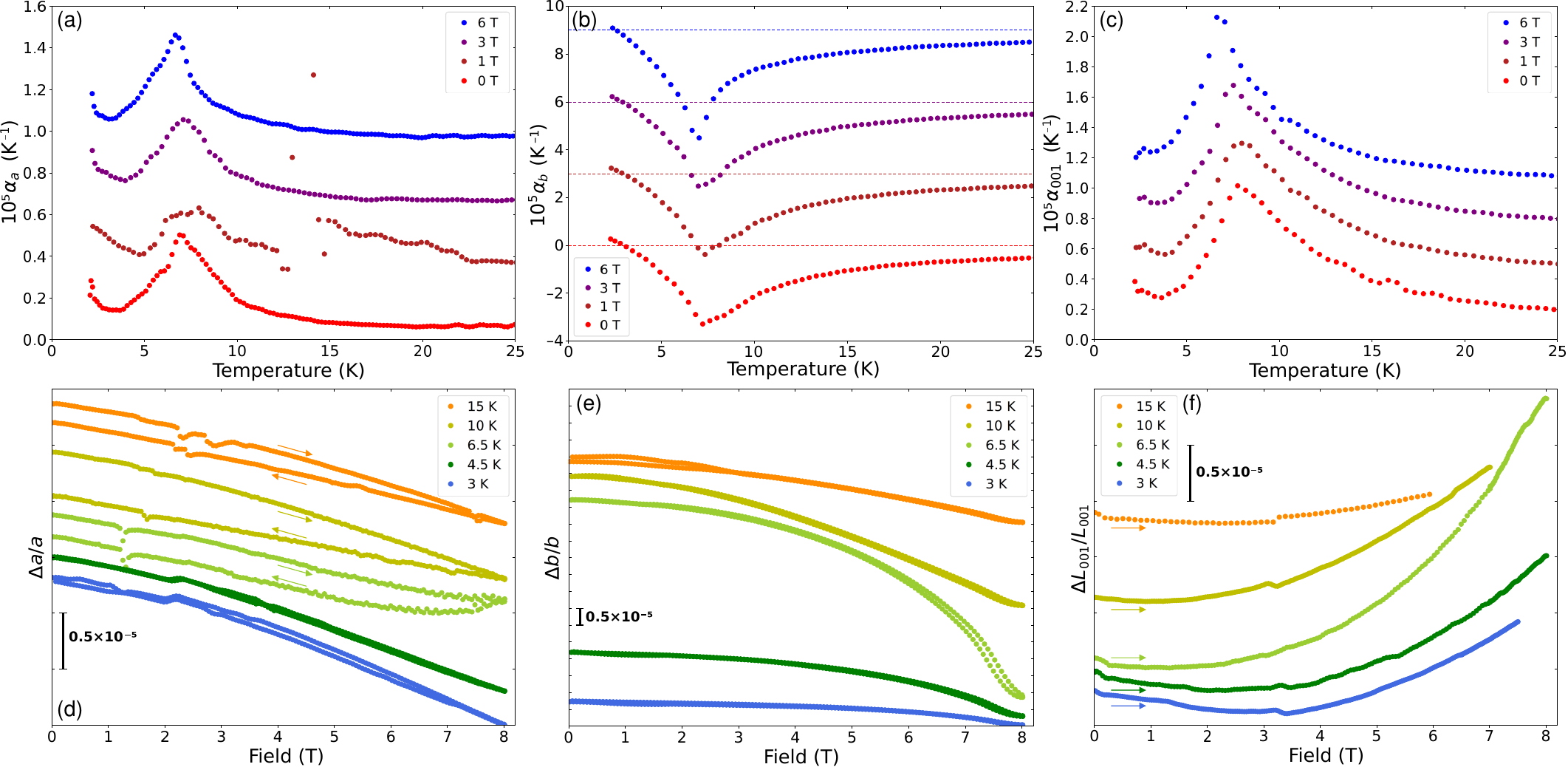}
  \caption{\label{striction}Linear thermal expansion and magnetostriction of rouaite.  (a-c) Linear thermal expansion coefficients $\alpha_i$ for the (a) $a$, (b) $b$, and (c) [0\,0\,1] directions as a function of temperature, for longitudinal fields.  In these panels, the datasets in field have been shifted vertically by multiples of $3\times 10^{-6}$\,K$^{-1}$ for visual clarity.  (d-f) Magnetostriction along the (d) $a$, (e) $b$, and (f) [0\,0\,1] directions, again for fields parallel to the expansion.}
\end{figure*}

We turn to dilatometry measurements to investigate this effect more sensitively. The relative length changes in the $a$, $b$, and [0\,0\,1] directions at zero field are plotted against temperature in Fig.~\ref{dil_axes}.  As found in the neutron refinements in Fig.~\ref{WombatExp}, the $b$ axis drives significant negative volume expansion above \TN.  An inflection point is found at the N\'eel transition for all three directions.  Here, we also see that the $a$ and [0\,0\,1] directions have opposite curvature above \TN, and that positive thermal expansion is restored along the $b$ axis below $\sim$3\,K. The dilatometric results are in very good agreement with those from neutron scattering, but the magnitude of the changes measured by dilatometry on our single crystal is roughly half that seen by diffraction on powder. This discrepancy is not fully understood, with one potential contribution being neglected stray capacitances.

From these thermal expansion data, it is possible to calculate the relative change in volume $\Delta V/V$ and the volumetric thermal expansion coefficient $\gamma = \frac{1}{V} \frac{\text{d}V}{\text{d}T}$, which are plotted for several fields in Figs.~\ref{dil_vol}(a) and \ref{dil_vol}(b), respectively.  Here we find that positive volumetric thermal expansion is restored on cooling below $\sim$4\,K.  The volume changes exhibit a strong kink at \TN, which manifests as a strongly asymmetric cusp in $\gamma$.  The shape of $\Delta V/V$ echoes that seen in Fig.~\ref{WombatExp}(e).

Dilatometry was also used to investigate the magnetostriction and the thermal expansion in field, through measurements in both transverse and longitudinal fields (expansion perpendicular or parallel to the applied field, respectively).  We present the longitudinal thermal expansion data in Figs.~\ref{striction}(a-c) for three orthogonal directions, plotted as the linear thermal expansion $\alpha$, where for instance $\alpha_a = \frac{1}{a} \frac{\text{d}a}{\text{d}T}$.  For all three directions, a strong cusp is observed at \TN.  In many cases there are hints of a weak kink or shoulder on the high-temperature side of \TN, presumably associated with short-range order, while for the $a$ and [0\,0\,1] directions, the thermal expansion reaches a minimum around 3--4\,K before increasing again at lower temperatures.  These upturns underlie the strong increase in the volume expansion below 3\,K in Fig.~\ref{dil_vol}(b).  

The magnetostriction is shown in Figs.~\ref{striction}(d-f).  At most temperatures, including well above \TN, this is slightly hysteretic, due to issues with temperature stability.  Just above 2\,T for $H\parallel a$ and 3\,T for $H\parallel [0\,0\,1]$, a step is observed which likely corresponds to the higher-field transition seen in the magnetization.  Interestingly, this may still be present at 10\,K for $H\parallel [0\,0\,1]$, which is above \TN.  Curiously, in the magnetostriction, [0\,0\,1] fields produce opposite curvature to the other two directions, leading to a minimum at finite field.

\begin{figure*}
  \includegraphics[width=\textwidth]{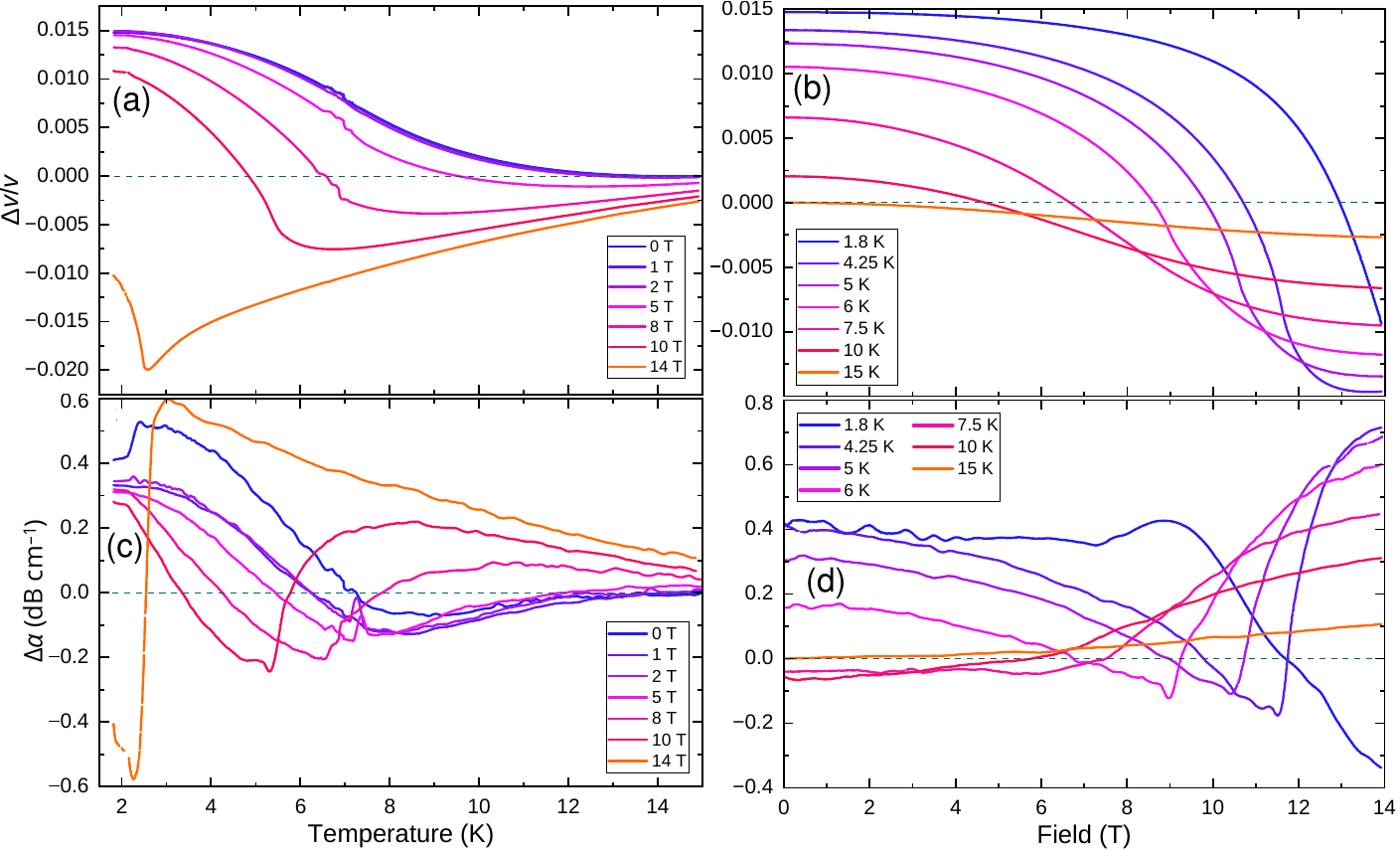}
  \caption{\label{ultra}Ultrasound results on rouaite, with $H$ and sound wave polarization and wavevector along $b$.  (a,b) Relative change in sound velocity as a function of (a) temperature and (b) magnetic field. (c,d) Relative change in attenuation as a function of (c) temperature and (d) magnetic field.  Due to the necessity of adjusting the measurement frequency slightly between runs, the temperature sweeps were put on a common vertical scale by shifting to match the field sweep at 15\,K and scaling to match the 1.8-K data, then verifying consistency among the datasets at intermediate temperatures and fields.}
\end{figure*}

\section{Acoustic Properties}

We investigated changes in the ultrasound velocity and ultrasound attenuation as a function of temperature and field $H\parallel b$ to gain further insight into the magnetic transitions.  Sound waves can interact with magnetic order either through (i) altering the single-ion anisotropy through the crystal electric field due to the (moving) ligands and spin-orbit coupling, or (ii) exchange striction, whereby moving the atoms alters the exchange pathways.  Single-ion anisotropy is absent in spin-$\frac{1}{2}$ systems, so only exchange striction is relevant in rouaite.  

In the sound velocity [Figs.~\ref{ultra}(a,b)], only the N\'eel transition is visible.  This manifests as an inflection point in the temperature and field dependence.  Given the large magnetoelastic coupling, it is not surprising that rouaite becomes more rigid on cooling below the transition --- the magnetic order contributes to the lattice stiffness.  Surprisingly, \TN\ is associated with a minimum in the attenuation [Figs.~\ref{ultra}(c,d)], --- no attenuation maximum caused by fluctuations around the transition is observed for this acoustic mode. Note that the total attenuation change in this temperature and magnetic field range is rather small, although the sound velocity change is rather large.

The metamagnetic transition is associated with a broad inflection point in the temperature-dependent attenuation.  However, it has no clear signatures in the ultrasound velocity or in field sweeps.  The relatively minor coupling of this transition to the lattice suggests that it is a reorganization of the existing order.

\section{Magnetic phase diagram}

\begin{figure*}
  \includegraphics[width=\textwidth]{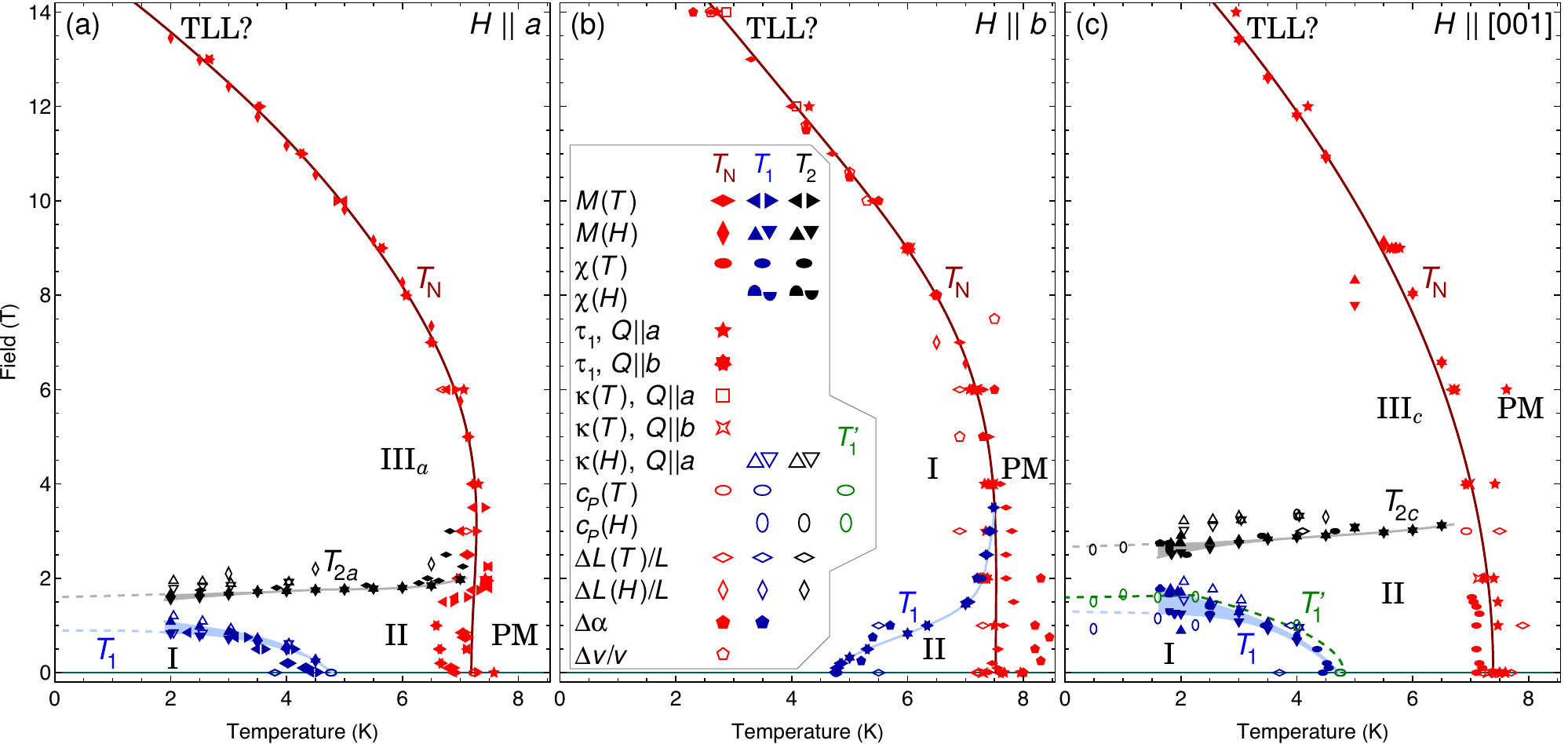}
  \caption{\label{PhasDiag}$H$--$T$ phase diagrams of rouaite for magnetic field along (a) $a$, (b) $b$, and (c) [0\,0\,1] (perpendicular to the $ab$ plane).  Uncertainties are smaller than the symbol size.}
\end{figure*}

$H$--$T$ phase diagrams for magnetic fields along $a$, $b$, and [0\,0\,1] (perpendicular to the $ab$ plane) based on the data described above are presented in Fig.~\ref{PhasDiag}.  Transitions are also clearly visible in the relaxation time in thermal transport measurements, and in some cases also in the thermal conductivity $\kappa$, and these have also been added to the phase diagrams; details of these results will be reported elsewhere.  As in several other measurements, transitions among the magnetically ordered phases were hysteretic in thermal transport.  In magnetization and ultrasound data it was possible to identify a maximum slope in field sweeps at higher temperatures, which would lead to an additional curve departing from \TN\ around 6-8\,T and reaching zero field around 14\--17\,K.  This corresponds approximately to the maximum in temperature sweeps, so this broad crossover presumably arises from an onset of short-range order.  As such, it is not plotted, and the region above \TN\ is paramagnetic (labelled PM), at least at low field.

For $H\parallel a$ and $H\parallel [0\,0\,1]$, the phase boundaries are not significantly different from those reported in Ref.~\onlinecite{Yuan2022}, but several matters are clarified.  The transition which begins at 4.72\,K at low field and falls to zero temperature around 1--2\,T exhibits strong hysteresis (shaded regions in Fig.~\ref{PhasDiag}), suggesting it to be first order.  We call this transition $T_1$, the phase below it I, and the phase at higher temperature II.  The transition around 2--3\,T, which we call $T_2$ and which bounds a high-field phase~III, also exhibits some hysteresis and is likely also first order.  The previous work presented these transitions as being isotropic and merging at low temperature with a shape suggestive of a quantum critical point.  In our data, however, these transitions occur at different fields in these two field orientations, are both first order, and remain well separated to low temperatures.  \TN\ is also suppressed to zero at higher field for $H\parallel [0\,0\,1]$.  That these two field orientations apparently differ by only a small overall rescaling of the field suggests that for fields perpendicular to the chains, there is only minor anisotropy.  We nonetheless label the $T_2$ transition and phase~III by field direction, since we cannot demonstrate that these evolve continuously into one another.  We will revisit this in the Discussion.

Our specific heat measurements in $H\parallel [0\,0\,1]$ suggest an additional transition which we have labelled $T_1^\prime$, just above $T_1$.  However, since this was not seen with any other technique, its existence remains tentative.  If present, this would suggest a sliver of an additional phase~I$^\prime$ roughly corresponding with the $T_1$ hysteresis range.  We would expect a similar sliver for $H\parallel a$, but how this transition would evolve with field $H\parallel b$ is unclear.  

\begin{figure*}
  \includegraphics[width=\textwidth]{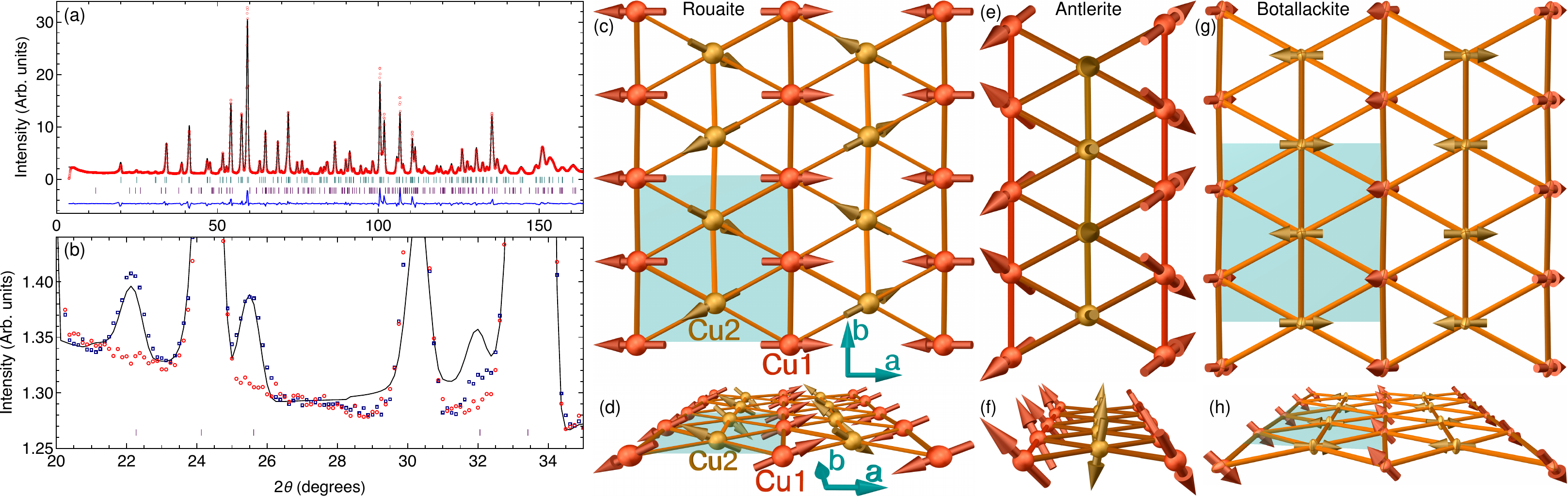}
  \caption{\label{Ground}Magnetic ground state of rouaite, \rouaD.  (a) Refinement of the crystal and magnetic structures at 3.2\,K based on data from Echidna. Data are red, the calculated profile is black, and the residual is blue, while turquoise and purple bars mark structural and magnetic reflections, respectively.  (b) View of the first few magnetic peaks, comparing 20- (red) and 3-K (blue) data collected on the Wombat high-intensity diffractometer. Purple bars mark magnetic reflections. (c) Refined low-temperature magnetic structure, based on the Echidna data in (a).  One Cu$^{2+}$ plane is shown, and the structural unit cell is shaded. (d) Side view of the magnetic structure. (e,f) Comparison against an isolated three-leg ladder in antlerite, Cu$_3$(SO$_4$)(OH)$_4$\,\cite{Kulbakov2022a}.  (g,h) Comparison against the published magnetic ground state in botallackite \botDBr\,\cite{Zhang2020}; the structural unit cell is shaded.}
\end{figure*}

The phase diagram for $H\parallel b$ is very different from the other two field orientations and from what was proposed previously\,\cite{Yuan2022}.  In particular, and in stark contrast to the other field orientations, phase~I appears to persist to high field, dominating the phase diagram. Phase~II is squeezed out to {\slshape high} temperature rather than low, and $T_2$ and phase~III are  either entirely absent or a continuous deformation of the low-field low-temperature phase.  The previously published phase diagram likely had too few measurements for $H\parallel b$ to resolve this behaviour.

$T_1$ and $T_2$ both exhibit greater hysteresis at lower temperatures, likely because the latent heat gets large compared to the temperature.  For $b$-axis fields, where the transition stays at higher temperatures, no significant hysteresis was observed.

\section{Low-temperature magnetic structure}

\begin{figure}[tb]
    \includegraphics[width=\columnwidth]{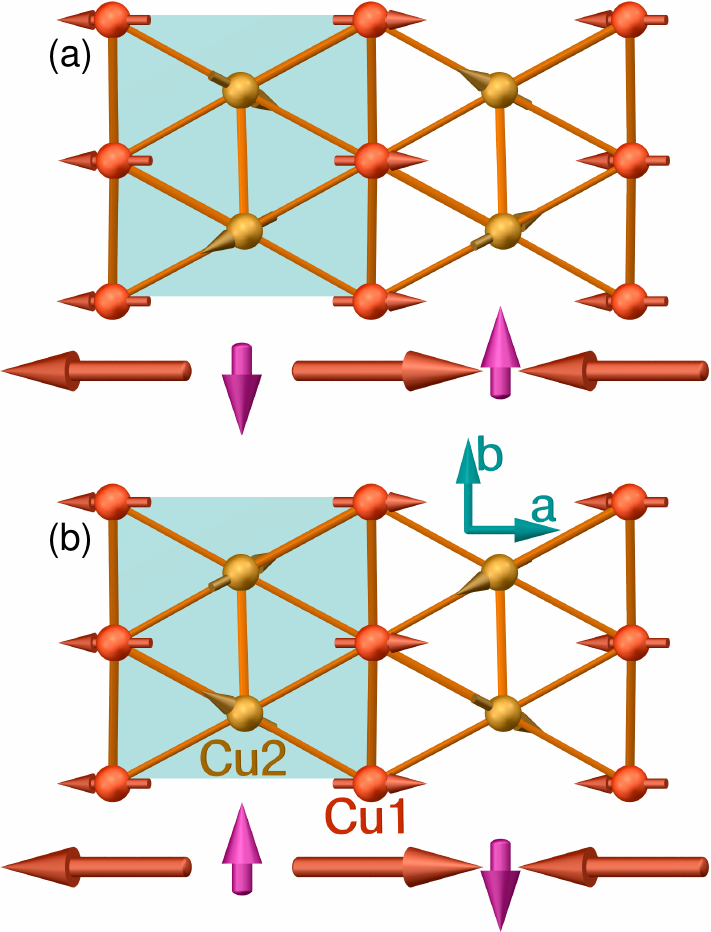}
    \caption{\label{cycloid}The two possible senses of rotation in the approximately cycloidal low-temperature phase~I.  (a) The version presented in Fig.~\ref{Ground}(c,d).  (b) A state with opposite circulation but identical neutron powder diffraction pattern.  The net moments on each chain are shown at the bottom of each panel.}
\end{figure}

A refinement of the crystal and magnetic structures in phase~I at 3.2\,K based on data from the Echidna diffractometer is shown in Fig.~\ref{Ground}(a).  An expanded view of a similar result from the Wombat diffractometer, comparing data collected at 3 and 20\,K, is shown in Fig.~\ref{Ground}(b).  Here the first few magnetic peaks are visible.  As depicted in Fig.~\ref{Ground}(c,d), our refinements indicate magnetic order closely related to that of botallackite \botDBr\,\cite{Zhang2020} [shown for reference in Fig.~\ref{Ground}(g,h)], with alternating ferromagnetic and antiferromagnetic chains.  The propagation vector is ($\frac12$\,0\,0) as in \botDBr, and this magnetic state is similar to that calculated for rouaite in Ref.~\onlinecite{Ruiz2006}.  Our results do not support earlier or later calculations suggesting antiferromagnetic interactions along both chains\,\cite{Linder1995,Yuan2022}.  As in antlerite [Fig.~\ref{Ground}(e,f)\,\cite{Kulbakov2022a}], the spins in rouaite lie neither parallel nor perpendicular to the plane of their Cu sublattice.  We also find evidence for canting along the $b$ axis for one of the two Cu sites, as in antlerite.  This canting angle has a crucial difference from that of antlerite, however:  in antlerite, the ferromagnetic chains exhibit an antiferromagnetic canting along $b$, whereas in rouaite this canting is {\slshape ferromagnetic}.  No such canting angle was reported in \botDBr\,\cite{Zhang2020}, nor was it captured in the calculations in Ref.~\onlinecite{Ruiz2006}.  Ferromagnetic canting means that each antiferromagnetic chain has a net moment along $b$, as shown in Fig.~\ref{cycloid}.  If we replace every chain with its net moment, the overall magnetic state will represent an elliptical commensurate cycloid propagating along $a$ with a 90$^\circ$ rotation of the net moment from one chain to the next.  This represents a symmetry which is broken in rouaite but not botallackite or antlerite.  

Cycloidal order is ordinarily not chiral, since the plane of the cycloid is a mirror plane; however, the monoclinic angle in rouaite breaks this mirror and the ferromagnetic moments are canted out of the $ab$ plane, so circulation about [0\,0\,1] and [0\,0\,$\overline{1}$] are inequivalent.  It is likely that one chirality inherits a higher stability from the $P2_1$ crystal structure, but our refinements cannot distinguish the two possible chiralities for this cycloid, and the two-fold screw axis in $P2_1$ does not have a handedness that would offer a hint.  We show both circulations in Fig.~\ref{cycloid}, but use the structure in Fig.~\ref{cycloid}(a) in our discussion and figures.  The cycloid's chirality could be determined experimentally by neutron spherical polarimetry, for instance, and a detailed theoretical calculation of the ground state may suggest which chirality is more stable.  However, the selection of a preferred chirality by the lattice may rely on weak interlayer interactions, in which case the energy difference may be small and the cycloidal order could potentially form magnetic twin domains.

Previous calculations of the magnetic interactions have indicated that the strongest interaction is $J_\text{1}$ along the ferromagnetic chain, with the antiferromagnetic chain's $J_\text{2}$ not far behind\,\cite{Ruiz2006}.  However, the significant expansion along $b$ with cooling, together with the Goodenough-Kanamori rules whereby Cu--O--Cu angles approaching 180$^\circ$ are optimal for antiferromagnetic interactions, suggest that the antiferromagnetic $J_\text{2}$ interactions are actually stronger, and they stretch the lattice to improve their own stability.  This expansion should thus have a very significant impact on theoretical calculations of the magnetic ground state.  

\section{Intermediate-temperature magnetic structure}

Having identified the low-temperature phase, we now turn to phase~II.  The extremely weak feature(s) separating phases I and II in the specific heat coupled with the lack of a change in power law suggests that it is a minor reorganization of the lower-temperature order, with all spins continuing to participate in a manner similar to that found at lower temperature.  The temperature dependence of the magnetization for in-plane fields exhibits sharp and significant jumps at this transition, with opposite jumps for $H\parallel a$ and $b$.  Together with the much weaker signatures in the magnetization for $H\parallel [0\,0\,1]$, this suggests a rotation of (parts of) the existing spin order within the $ab$ plane.  

\begin{figure}[tb]
    \includegraphics[width=\columnwidth]{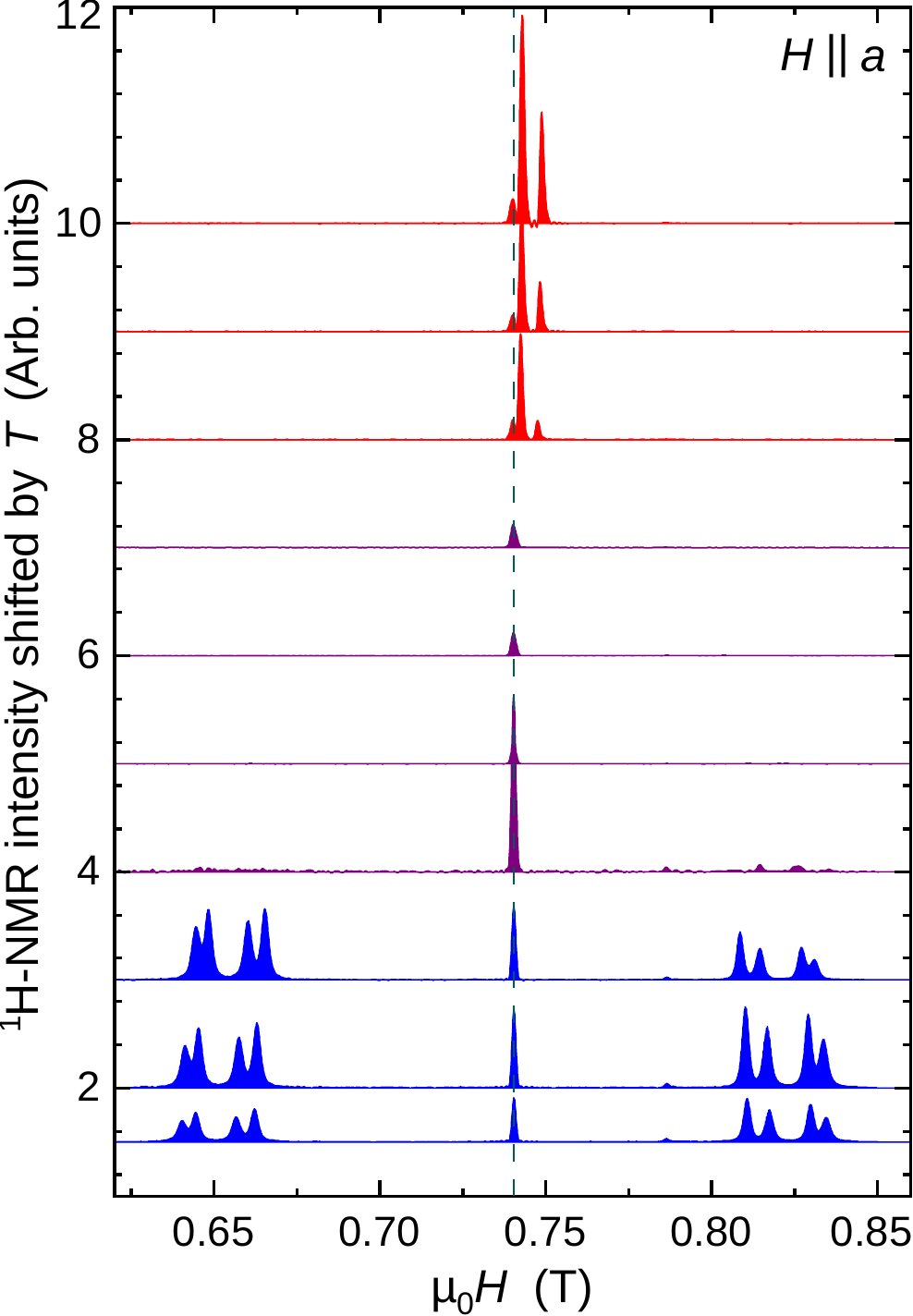}
    \caption{\label{NMR1}$^1$H-NMR spectra for fields along $a$ at various temperatures (in K) corresponding to their baselines.  The vertical dashed line indicates a nonmagnetic contribution from the sample holder, and colours correspond to the magnetic phases.  The vertical scale can vary among the spectra due to tuning of the measurement circuit, but all purple spectra have been expanded significantly.}
\end{figure}

\begin{figure}[tb]
    \includegraphics[width=\columnwidth]{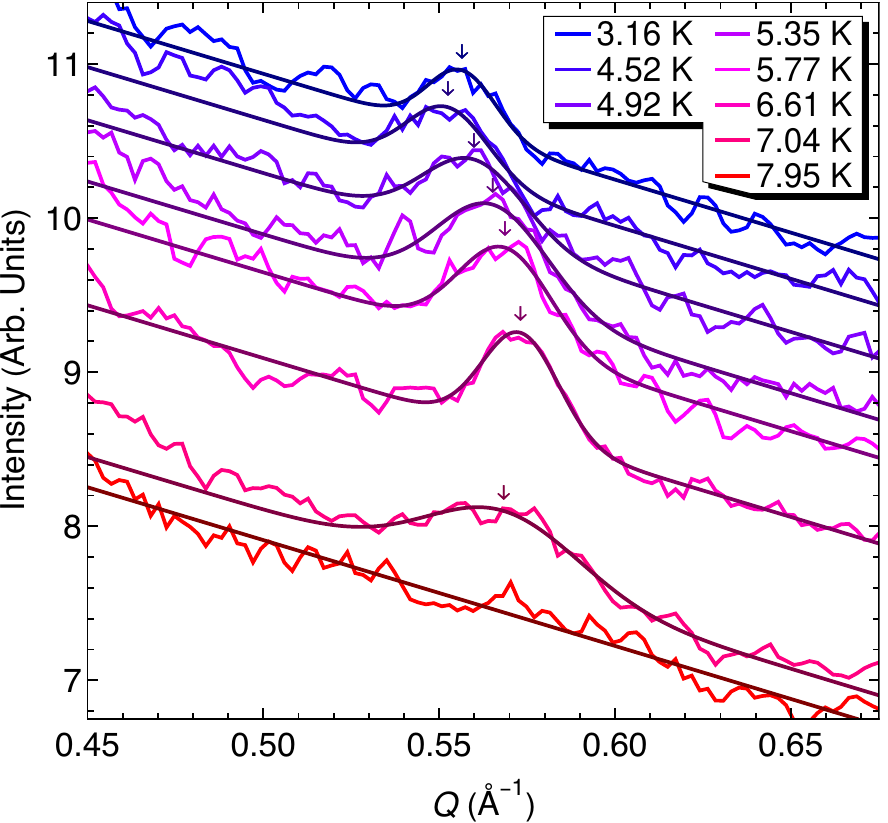}
    \caption{\label{Echidna_shift}Position of the magnetic peak corresponding to ($\frac12$\,0\,0) in phase~I, as a function of temperature through phase~II.  Peak positions are marked for each Gaussian fit.  To improve visual clarity, a 3-point moving average has been applied to the datasets and consecutive datasets have been shifted vertically by 0.4.}
\end{figure}

Proton NMR spectra measured on a 65\%-deuterated crystal in an applied magnetic field of 0.7\,T parallel to the $a$ axis are shown in Fig.~\ref{NMR1}.  Other field directions (not shown) behaved similarly.  At low temperatures (blue), peaks corresponding to multiple proton sites are observed, with splitting and temperature-dependent shifts.  This behaviour indicates that phase~I represents bulk, commensurate magnetic order which is static on the timescale of the NMR measurement.  Above \TN, in the paramagnetic state (red spectra), two peaks corresponding to the Cu1 and Cu2 sites are also visible, with only slight shifts relative to protons in nonmagnetic environments in the sample holder (dashed line).  However, at intermediate temperatures (purple) the peaks arising from the sample are all but absent, despite the intensity of these spectra having been enhanced relative to the others.  Vestiges of the peaks may potentially still persist, likely due to a small fraction of the sample in which the low-temperature phase is somehow pinned, but this cannot be a significant fraction of the sample.  Such a near-complete wiping out of the spectra is a fingerprint of incommensurately ordered phases, where the field strength and direction is different at every proton site.  

Figure \ref{Echidna_shift} shows the temperature evolution of the phase-I ($\frac12$\,0\,0) magnetic Bragg peak from 3 to 8\,K through phase~II, in neutron powder diffraction data collected on Echidna. 
Each dataset has been fit in $2\theta$ to a Gaussian, an offset, plus the sloping background determined at 7.95\,K in the paramagnetic phase, and the centre of each Gaussian has been marked on the figure.  There is a clear evolution of the peak position with temperature, again indicating that phase~II is not commensurate.  We note that this ($\frac12$\,0\,0) peak does not have a component sensitive to the $b$ direction where strong negative thermal expansion is observed, and in any case these changes in $Q$ at the $\sim$5\%\ level below \TN\ are three orders of magnitude weaker than the changes in the $b$ axis above \TN, excluding thermal expansion as a potential explanation.  This movement of the magnetic reflection is only possible if it is moving in momentum space, implying that the order must be incommensurate.

\begin{figure}[tb]
    \includegraphics[width=\columnwidth]{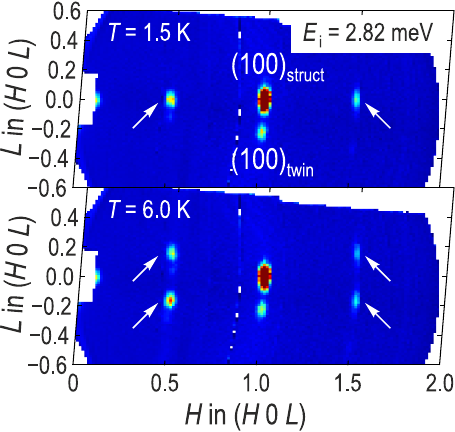}
    \caption{\label{LET}Elastic scan of the $(H0L)$ plane centred around the structural (1\,0\,0\,) peak, based on data collected at LET with incoming energy 2.82\,meV.  The ($\frac12$\,0\,0) and ($\frac32$\,0\,0) magnetic reflections at 1.5\,K, marked with arrows in the upper panel, split on warming into the intermediate-temperature phase (lower panel).  These magnetic reflections split approximately along $c$ (vertical), not $L$ ($\sim$5$^\circ$ off vertical).  Roughly 9\%\ of this mosaic sample was rotated 180$^\circ$ within the $ab$ plane, leading to different orientations of the $c$ axis and the in-plane reciprocal lattice vectors and giving rise to additional reflections; this fraction of the sample produces the structural peak marked ``(10\,0\,)$_\text{twin}$''.}
\end{figure}

\begin{figure}
  \includegraphics[width=\columnwidth]{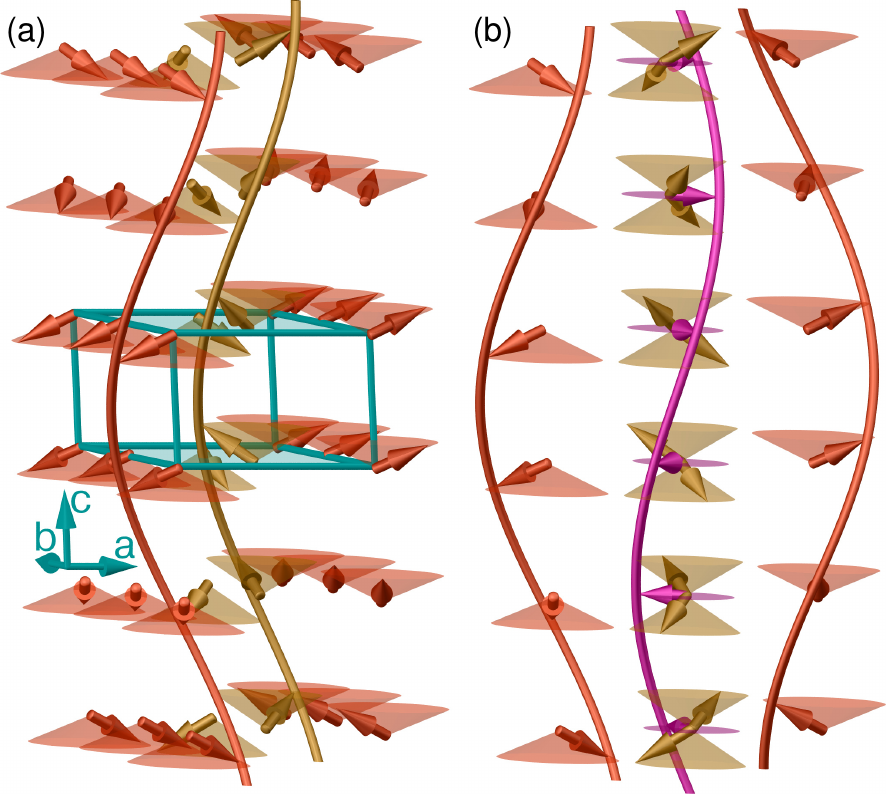}
  \caption{\label{helix}Intermediate-temperature incommensurate magnetically ordered phase~II.  (a) Several unit cells of rouaite, with the $c$ axis compressed by a factor of 2.  The spins trace out cones as they rotate along $c$.  The helical trajectories of one Cu1 spin and one Cu2 spin are traced out. (b) Trajectories of the spins on three consecutive chains, including the net moment on the Cu2 chain in magenta, viewed along the chains ($b$ axis).}
\end{figure}

It is possible to glean further details about the nature of the magnetic order in phase~II from time-of-flight neutron data collected on the LET spectrometer at the ISIS neutron source, Didcot, UK.  Figure \ref{LET} shows elastic scans of the $(H0L)$ plane at 1.5 and 6.0\,K, integrated over $\pm$0.05\,meV in energy and $\pm$0.05 in $K$, using an incoming energy of 2.82\,meV.  Deep in the low-temperature magnetically ordered state at 1.5\,K, spots corresponding to the phase-I order are observed at ($\frac12$\,0\,0) and equivalent positions.  In phase~II at 6.0\,K these are split perpendicular to ($\frac12$\,0\,0), indicating an incommensurate stacking of the layers.  Based on the magnetization and specific heat results, this stacking must be approximately helical.  At 6.0\,K, the pitch of the helix is approximately 6 unit cells.  It is important to note that since there is no surviving intensity at the commensurate position, the spins on both Cu sites must participate in the helical modulation.  Interestingly, the splitting of these magnetic peaks is not along $L$ --- it is closer to the $c$ direction, which would skew the helical modulation by several degrees.  

\begin{figure}[t]
    \includegraphics[width=\columnwidth]{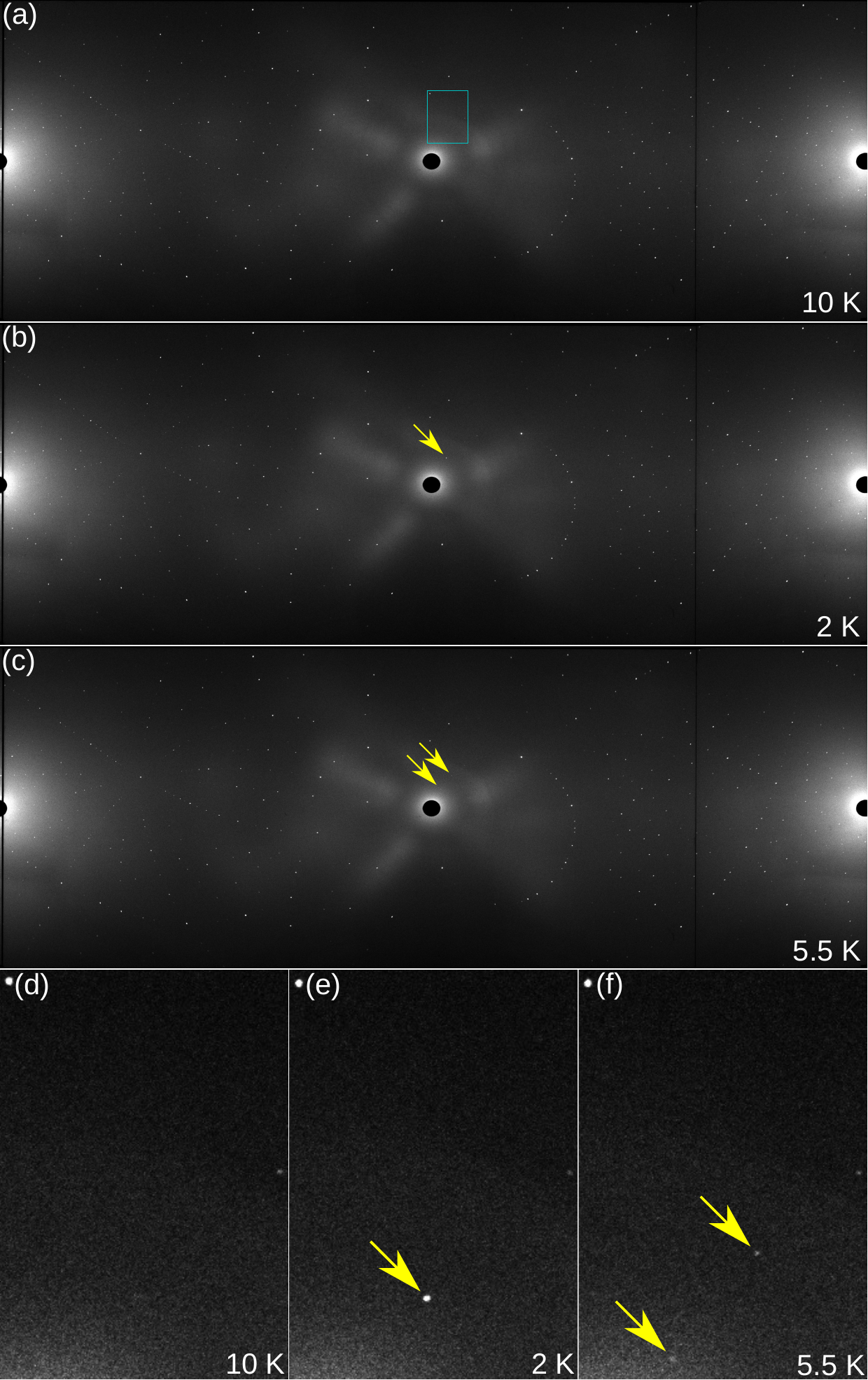}
    \caption{\label{Koala}Neutron Laue diffraction patterns on \rouaH\ at (a) 10\,K in the paramagnetic state, (b) 2\,K in the low-temperature state, and (c) 5.5\,K in the incommensurate phase.  Arrows mark the strongest magnetic reflections for this sample angle;  in (d) this is ($-\frac{1}{2}$\,0\,1).  Expanded views of the region corresponding to the blue box in (a) are shown at higher contrast for (d) 10\,K, (e) 2\,K, and (f) 5.5\,K.}
\end{figure}

A schematic magnetic structure of phase~II is shown in Fig.~\ref{helix}(a).  Each spin rotates on a cone as we move from layer to layer, and the tip of each spin traces out a helix which is slightly skewed by the monoclinic angle.  Spins representative of each chain are replotted in Fig.~\ref{helix}(b), where the net moments on the antiferromagnetic Cu2 chains are shown in magenta.  As for the commensurate cycloidal state at low temperature for which we show both possible chiralities in Fig.~\ref{cycloid}, we cannot distinguish either the circulation of the cycloid or the helicity of the stacking in phase~II, so there are four possible combinations of which we show only one.  As in phase~I, these are most likely not degenerate --- the $P2_1$ crystal structure should again prefer one of the four combinations, although the energy differences among them may be small.  

Neutron Laue diffraction performed on the Koala2 beamline at ANSTO, Australia, confirms this result, as shown in Fig.~\ref{Koala}.  In the example images presented, aside from increased intensity on several structural peaks, a strong magnetic peak appears at 2\,K [Fig.~\ref{Koala}(b,e)] which is not present at 10\,K [Fig.~\ref{Koala}(a,d)].  At 5.5\,K this peak, which we index as ($-\frac{\text{1}}{\text{2}}$\,0\,1), is split approximately along the $L$ direction [Fig.~\ref{Koala}(c,f)].  However, the peaks are not located symmetrically about ($-\frac{\text{1}}{\text{2}}$\,0\,1), consistent with a splitting which is not along $L$.  A crystal structure refinement of the 10-K structural data is provided in Tables \ref{NPDsummary} and \ref{NXDKoalaAtom} and Fig.~\ref{Koala10KF2F2} in Appendix~\ref{appA}, and crystallographic information files (CIFs) are available as arXiv ancillary files online, see Appendix~\ref{supp}.

\section{Discussion}

The picture that emerges from these results is that of a Cu sublattice with botallackite-like magnetic order, but in which weak interactions, such as presumably the interlayer exchange coupling, are able to shift the balance and drive the magnetic order through first-order phase transitions into additional magnetic phases.  

In phase~I, the magnetic susceptibility $M/H$ is significantly lower for $H\parallel a$ than for the other directions, as can be seen for instance in Figs.~\ref{MvsH}(a) and \ref{MvsH}(d).  As shown in Fig.~\ref{Ground}(c), the magnetic order in this phase has spins predominantly pointing along $a$, giving it a very limited ability to adapt to applied fields along this direction.  This low-temperature spin arrangement rotates from plane to plane in the helical state (phase~II), putting $a$ and $b$ on an even footing, consistent with their opposite jumps.  This rotation is perpendicular to [0\,0\,1], so it is unsurprising that the transitions are difficult to observe in that direction.  This transition requires completely reorienting entire slabs of spins to reach a state of nearly identical entropy, explaining the very small but nonzero latent heat.  Rotating a slab of spins within the $ab$ plane to generate the helical phase is presumably driven by interlayer exchange interactions, but these exchange pathways are long and involve weak hydrogen bonds (i.e., Cu---O---H\,$\cdots$O---N---O---Cu).  This suggests that the energy barrier for rotation is very low.  

The interlayer exchange pathway is significantly simpler in botallackite, where the anion is a simple halide, but no transition to an incommensurate structure has been observed in the chloride, bromide or iodide\,\cite{Zheng2005,Zheng2009,Zhang2020,Xiao2023}.  We note, however, that magnetization measurements on the chloride and iodide were only performed on powder, and in rouaite powder $T_1$ is not observable\,\cite{Linder1995,Drillon1995,Kikuchi2018}.  The helical phase likely evidences a delicate balance of interactions within the plane\,---\,a balance which can be tipped by weak interlayer exchange or perhaps Dzyaloshinskii-Moriya interactions.  It is not clear why the field strength required to reach this phase is similar for fields along $a$ and [0\,0\,1], but the fact that a higher field is required to {\slshape destroy} the incommensurate phase in [0\,0\,1] fields is unsurprising --- when spins order helically, applying magnetic fields perpendicular to the plane of the spins cants the spins to make the order slightly conical, while fields in the plane of the spins destroy the order.  This effect has been used to detwin or partially detwin helical magnetism in ZrCr$_2$Se$_4$\,\cite{Inosov2020}, SrFeO$_3$\,\cite{Ishiwata2020}, and Sr$_3$Fe$_2$O$_7$\,\cite{Andriushin2023}, among other materials.  The completely different behaviour in $b$-axis fields likely stems from the relative stability of the commensurate state for this field orientation.

The narrow sliver of phase~I$^\prime$, if it exists, would presumably correspond to a state in which either the Cu1 or the Cu2 site is helical, but the other is commensurate, in the magnetic analogue of the misfit layered compounds.  This would suggest that one site drives the helical order, dragging the other along with it.  A similar scenario was reported in antlerite, in which such a magnetic misfit phase was found between 5.00 and 5.30\,K at zero field\,\cite{Kulbakov2022b}.  In rouaite, this tentative phase only exists in a 0.06-K-wide window in zero field, and accessing the phase would require carefully oscillating field or temperature while approaching it, to avoid the hysteresis.  Verifying its presence would be possible, and interesting given that this would be only the third reported magnetic misfit phase after those in antlerite\,\cite{Kulbakov2022b} and Co$_3$TeO$_6$\,\cite{Ivanov2012,Wang2013,Lee2017}, but this verification would not be straightforward.  

Before addressing phase~III, we turn to the phase above \TN.  At high field, the ferromagnetic chains would polarize and the ferro- and antiferromagnetic chains would likely decouple, as has been reported in antlerite\,\cite{Kulbakov2022a} and botallackite Cu$_2$(OH)$_3$Cl\,\cite{Xiao2023}.  Previous work on rouaite points to transitions of this nature around 15\,T at low temperature, into a $\frac{2}{3}$ plateau phase corresponding to fully polarized ferromagnetic chains\,\cite{Yuan2022}.  This occurs where the transition we label as \TN\ is suppressed to zero temperature.  Such a phase, in which the antiferromagnetic chains can no longer readily talk to each other through saturated field-polarized ferromagnetic chains, is a prime candidate for an exotic field-induced Tomonaga-Luttinger liquid\,\cite{Kulbakov2022a}, and is accordingly labelled ``TLL?'' in Fig.~\ref{PhasDiag}.  It does not represent long-range order, and is not separated from the low-field paramagnetic state by a phase transition.  In the case of rouaite, unlike in botallackite or antlerite, any vestiges of the low-field canting pattern would lead to a small net moment locally on each antiferromagnetic Cu2 chain, and dipolar interactions among the net moments on the short-range antiferromagnetic clusters may allow these chains to order at sufficiently low temperature.  

\begin{figure}[tb]
    \includegraphics[width=\columnwidth]{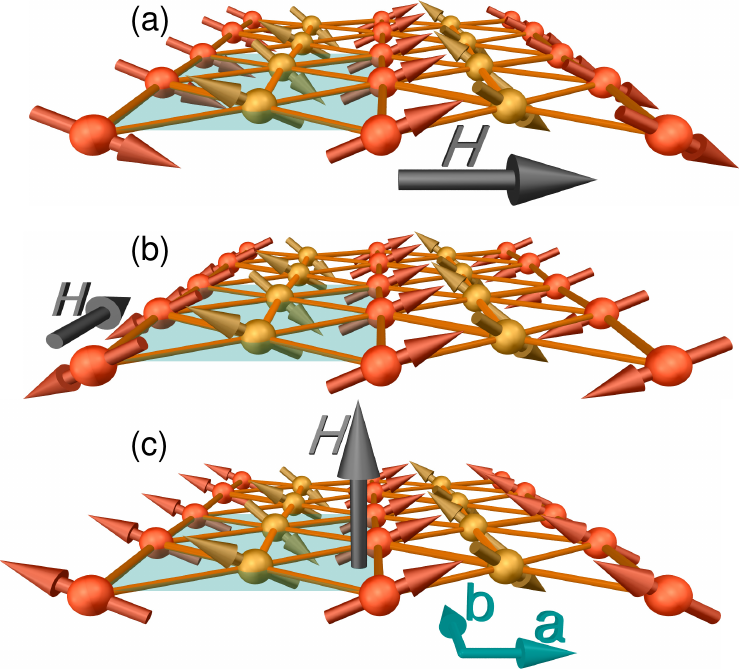}
    \caption{\label{III}Proposed effect of field on the single Cu plane shown in Fig.~\ref{Ground}(d), and inferred identity of phase~III.  (a) A field along $a$ is likely to flip the $a$ component of every second ferromagnetic chain.  (b) A field along $b$ will not produce a discontinuous change on the ferromagnetic chains.  At a higher field, it may eventually flip the net moment on alternate antiferromagnetic chains, possibly discontinuously (not shown).  (c) A field along [0\,0\,1] should flip the $c$-axis component of every second ferromagnetic chain.  The field will also slightly modify all canting angles, which we do not show here.}
\end{figure}

Phase~III, found over a broad field range for fields applied along $a$ and [0\,0\,1], cannot be conclusively identified based on our data, but we can speculate about its identity.  In phase~I and phase~II at zero field, the ferromagnetic chains have alternating net magnetic moment along both $a$ and [0\,0\,1], as can be seen in Fig.~\ref{Ground}(c,d).  Once a field is applied along one of these directions, half of the ferromagnetic chains are at higher energy than the others.  At fields exceeding the \TN\ line, the ferromagnetic chains are fully polarized and no longer alternate along the field direction.  This implies that the disfavoured ferromagnetic chains must flip at some intermediate field, likely in a discontinuous manner.  $T_2$ presumably corresponds to this transition.  Figure~\ref{III}(a) and (c) show what this will lead to for fields applied along $a$ and [0\,0\,1], respectively, for the single Cu layer shown in Fig.~\ref{Ground}(d).  For fields along $b$ [Fig.~\ref{III}(b)], the ferromagnetic chains can adapt continuously to the applied field and no such transition is expected, although an analogous transition on the antiferromagnetic Cu2 chain would be possible at much higher field.  

In the low-temperature limit, the angular dependence of $T_2$ may trace out a tilted ellipse or dumbbell as a function of angle in the $ac$ plane --- fields perpendicular to the spin orientation would polarize the Cu1 chains easily, leading to a minimum --- but it could also diverge for some field direction.  This will hinge on the strength of the interactions that prefer (1\,0\,1) spin orientation over ($\overline{1}$\,0\,1) in phase~I, which may be the weak interlayer interactions, and on the interactions that lead to helical order in phase~II.  We also note that a field applied along $a$ (approximately perpendicular to the helical propagation vector) would destroy the helical order.  Since it is not clear what selects the spin orientation or whether it is possible to rotate the field continuously from $a$ to [0\,0\,1] without encountering a phase transition, we cannot say whether phase~III would be the same for both $a$ and [0\,0\,1] fields.  These are therefore labelled with subscripts in the phase diagrams.  Phase~III$_a$ must be commensurate, but whether phase~III$_c$ remains helical will require further investigation.  The similar shape of the $T_2$ transition in the phase diagrams for these field orientations leads us to suspect that both are commensurate.  

While no analogue of the incommensurate phase~II has been reported in botallackite, a transition consistent with $T_2$ has been reported.  In polycrystalline botallackite \botHCl, this has been reported at 3\,T\,\cite{Zheng2009,Xiao2023}; it was observed at roughly 5\,T in single-crystalline \botHBr\ in powder\,\cite{Zheng2009} and for fields along $a$ and [0\,0\,1]\,\cite{Zhao2019} (again slightly higher for [0\,0\,1]); and in the iodide it occurs around 7\,T in powder\,\cite{Zheng2009}.  In the chloride, this field-induced transition was attributed to the antialigned ferromagnetic chains abruptly aligning\,\cite{Xiao2023} and the higher-field phase was labelled as ferromagnetic on the phase diagram.  However, this transition occurred at only around $\frac{1}{3}$ of the field required to reach the high-field plateau, implying that a significant canting angle must still be present at this spin-flop transition.  The $T_2$-like metamagnetic transition in the botallackites presumably has the same origin as in rouaite, most likely an abrupt change in the canting angle toward the field of every second ferromagnetic chain.

The phase diagram for $H\parallel b$ in rouaite is completely different from the other two directions, with the low-temperature phase being stabilized over the incommensurate phase instead of destabilized, and no transitions are seen with field at low temperature.  Phase~I can respond to $b$-axis fields by gradually and continuously polarizing along $b$, and since the ferromagnetic chains do not alternate along this direction, there is no $T_2$.  The spins on the antiferromagnetic chains do alternate along this direction, but with a much smaller net moment, so there may indeed be a transition but only a very weak one.  Given the structural evidence that the antiferromagnetic interactions along $b$ are very strong, any such transition would most likely occur at much higher field. 

The incommensurate phase~II is destabilized for $H\parallel b$ because the applied field is in the plane of the spins, so the phase diagram for this direction matches our expectations for phases I--III.  However, this raises two interesting points.  First, a field along [0\,0\,1] should also destabilize the helical order, presumably at a similar field, but $T_1$ for $b$-axis fields and $T_2$ for [0\,0\,1] fields are only similar where they merge with \TN, and their trajectories in the $H$--$T$ phase diagrams are nearly orthogonal.  It remains unclear why this is.  Second, the $T_1$, $T_2$, and tentative $T_1^\prime$ transitions which are present for field orientations perpendicular to the chains presumably must pinch off somehow as the field is rotated away from $b$.  It could be very interesting to find and investigate the angular dependence of these transitions.

In the phase diagrams in Fig.~\ref{PhasDiag}, several points seem not to match with the points around them, and in some areas there is considerable scatter.  The transition temperatures and fields often appeared to increase somewhat the longer the sample was kept cold.  While we cannot fully exclude artefacts from the apparatus, this would suggest that cycling the sample thermally and in magnetic field at low temperatures helps the structure relax.  The strong magnetoelastic coupling stretches the crystal along $b$ as seen in Fig.~\ref{WombatExp}, and our temperature sweeps typically extend up into the region where this turns on.  This stretching stabilizes exchange interactions at the expense of the Coulomb interactions holding the crystal together, and would be expected to stabilize the magnetic order.  A drift with cycling would be evidence of extremely high tunability of the magnetic order, and would suggest that hydrostatic and uniaxial pressure are likely to have extremely strong effects on this system.  

Finally, we note that the magnetic phases we identify in rouaite differ markedly from the resonating-valence-bond and weak zigzag-stripe phases proposed in Ref.~\onlinecite{Yuan2022}, which would potentially be in proximity to a quantum spin-liquid phase.  Hydrogen positions in that reference were not refined from the neutron data, so the starting point for theory calculations was an x-ray crystal structure refinement at 300\,K and hydrogen positions taken from density functional theory.  In antlerite, we found that the calculated magnetic structure was exquisitely sensitive to details of the interladder hydrogen positions, with the previously-published structure returning a ground state which was not only incorrect, but was not among four possible ground states ultimately found in proximity to the final exchange parameters\,\cite{Kulbakov2022a}.  The use of a room-temperature rouaite crystal structure for the calculations in Ref.~\onlinecite{Yuan2022} will also miss the significant deformation of the unit cell through strong magnetoelastic coupling at low temperature.  This will have a large effect on the strongest exchange pathways.  The ground state calculated in Ref.~\onlinecite{Yuan2022} most likely arose from these or similar subtle issues with the crystal structure, combined with the apparent high sensitivity of the magnetism to structural tuning.  Reference~\onlinecite{Yuan2022} did not identify magnetic intensity in their neutron diffraction data, most likely due to a high background of incoherent scattering from protons, attributable to the significantly lower deuteration level achieved.  This prevented experimental identification or refinement of the magnetic phases, which would have offered an important check on theory.  

\section{Conclusion}

Our magnetic phase diagrams for rouaite clarify the interesting behaviour in the low-field regime.  We identify the low-temperature magnetic state at low fields as being closely reminiscent of that of botallackite \botHBr, suggesting that rouaite may represent a second potential platform for probing spinon-magnon mixing, although this remains to be confirmed and there are some differences in the magnetic structure.  The large crystals reported here will enable inelastic neutron scattering to probe the excitation spectrum in search of this mixing.  The incommensurate phase at higher temperature suggests that interlayer exchange interactions, which should be extremely weak given the tortuous exchange pathways involved, nevertheless play a crucial role in stabilizing the magnetic order and are capable of driving a first-order magnetic phase transition. That this transition is not observed in the botallackites suggests that chemically tuning interlayer interactions will be a productive research direction in this family.  A transition at higher field, which is present in the botallackites is likely associated with the reversal of a canting angle on half of the ferromagnetic chains.  The evidence for strong magnetoelastic coupling, a delicate balance among interactions, and a central role for very weak interactions in tipping the balance and selecting a ground state make the tuning of rouaite through uniaxial or hydrostatic pressure particularly promising.  

Rouaite may also offer a unique opportunity to study and manipulate antiferromagnetic helical order.  The strong interactions in antiferromagnets ordinarily prevent helical order, while the lack of a net moment makes it difficult to manipulate the order with a laboratory-scale applied field\,\cite{Sato2020}.  The known helical antiferromagnets most commonly have near-ferromagnetic alignment within a layer together with antiferromagnetic layer stacking\,\cite{Kim2014,Sokolov2019,Jin2019,Lass2020}.  Materials in which consecutive spins are nearly antialigned are exceedingly scarce, with two of the best examples being FeP\,\cite{Felcher1971,Sukhanov2022} and CrAs\,\cite{Boller1971,Selte1971,Keller2015}.  In rouaite, the Cu2 chains have a primarily antiferromagnetic spin alignment but participate in both cycloidal and helical order, and the small net moment on the Cu2 chains may enable control through field.  

\section*{Data Availability}

Samples and data are available upon reasonable request from D.\ C.\ Peets or D.\ S.\ Inosov.  Data collected at ISIS are available through Ref.~\onlinecite{LET-RB2320001}.

\begin{acknowledgments}
This project was funded by the Deutsche Forschungsgemeinschaft (DFG, German Research Foundation) through individual grant PE~3318/2-1 (Project No.\ 452541981); through projects B03, C01, C03, and C10 of the Collaborative Research Center SFB~1143 (Project No.\ 247310070); and through the W\"urzburg-Dresden Cluster of Excellence on Complexity and Topology in Quantum Materials\,---\,\textit{ct.qmat} (EXC~2147, Project No.\ 390858490). The PPMS at TUBAF was funded through DFG Project No.\ 422219907.  The authors acknowledge the support of the Australian Centre for Neutron Scattering, Australian Nuclear Science and Technology Organisation, in providing neutron research facilities used in this work. This work is based in part on experiments performed at the Swiss spallation neutron source SINQ, Paul Scherrer Institute, Villigen, Switzerland. We gratefully acknowledge the UK Science and Technology Facilities Council (STFC) for access to neutron beamtime at ISIS and for the provision of sample deuteration facilities\,\cite{LET-RB2320001}. We acknowledge support of the HLD at HZDR, a member of the European Magnetic Field Laboratory (EMFL).
\end{acknowledgments}

\appendix
\section{Crystal Structure Refinement Details\label{appA}}

\begin{figure}[tb]
  \includegraphics[width=\columnwidth]{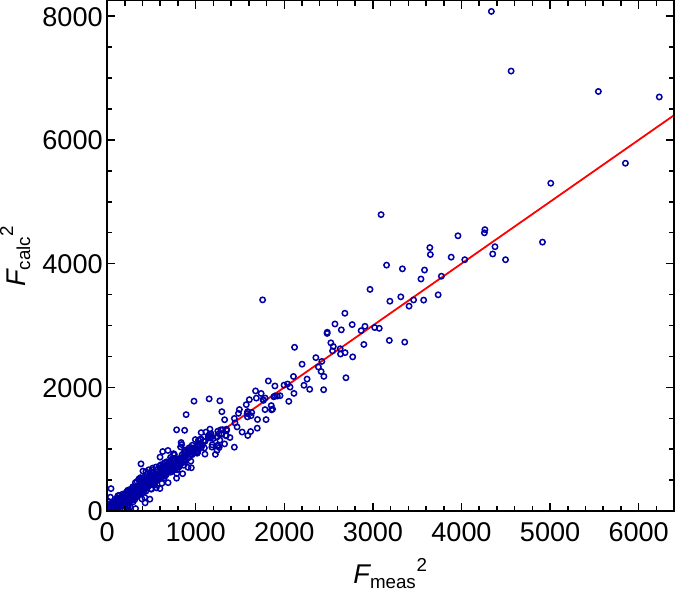}
  \caption{\label{Koala10KF2F2}Relationship between calculated and measured $F^2$ values for our refinement of Koala2 neutron Laue data at 10\,K, demonstrating the quality of the refinement.}
\end{figure}

This appendix contains details of our crystal structure refinements, which are summarized in Tab.~\ref{NPDsummary}.  Tables~\ref{NPDEch1} and \ref{NPDHRPT1} report the refined atomic positions in \rouaD\ based on powder data collected at 20\,K on Echidna, ANSTO, Australia, and at 10\,K on HRPT, PSI, Switzerland, respectively.  Our refined atomic positions in \rouaH\ based on neutron Laue data collected on Koala2, ANSTO, Australia, are listed in Tab.~\ref{NXDKoalaAtom}.  CIF files describing these refinements are provided in the ancillary files as part of this arXiv submission, see Appendix~\ref{supp}.  A plot of $F_\text{calc}^2$ {\itshape vs.}\ $F_\text{meas}^2$ for the neutron Laue refinement is provided in Fig.~\ref{Koala10KF2F2}, to indicate the quality of the refinement.


\begin{table}[b]
  \caption{\label{NPDsummary}Summary of crystal structure refinements of rouaite from neutron diffraction.  The lattice parameters for the Koala2 refinement were fixed at the data reduction stage, since neutron Laue diffraction is not sensitive to absolute lattice parameters.  The number of reflections refined in the Echidna data includes the $\frac{\lambda}{2}$ peaks.}
  \begin{tabular}{lrrr}\hline\hline
    & Echidna & HRPT & Koala2\\ \hline
    Type & Powder & Powder & Laue\\
    Temperature (K) & 20 & 10 & 10\\
    Space group & $P2_1$ (\#\ 4)& $P2_1$ (\#\ 4)& $P2_1$ (\#\ 4)\\
    $a$ (\AA) & 5.58683(3) & 5.58639(5) & 5.6005\\
    $b$ (\AA) & 6.06388(3) & 6.06448(5) & 6.0797\\
    $c$ (\AA) & 6.91127(5) & 6.91187(9) & 6.9317\\
    $\beta (^\circ)$ & 94.9011(5) & 94.9025(10) & 94.619\\
    $V$ (\AA$^3$) & 233.283(2) & 233.308(4) & 235.253\\
    $Z$ & 2 & 2 & 2\\
    Density (g cm$^{-3}$) & 3.45813(4) & 3.45776(6) & 3.4292\\
    Reflections & 1306 & 477 & 1405\\
    Reflections $>2\sigma$ &  &  & 994\\
    $2\theta$ range ($^\circ$) & 3.86--163.76 & 2.50--164.85 & \\
    Index ranges & & & $-6\leq h\leq 9$\\
    & & & $0\leq k\leq 11$\\
    & & & $0\leq l\leq 13$\\
    $F(000)$ &  &  & 96.794\\
    $R$ & 3.61\,\% & 5.21\,\% & 6.27\,\%\\
    $wR$ & 4.30\,\% & 6.76\,\% & 6.59\,\%\\\hline\hline
  \end{tabular}
\end{table}

\begin{table}[tb]
  \caption{\label{NPDEch1}Refined atomic positions in \rouaD\ from neutron powder diffraction using 2.440-\AA\ neutrons on Echidna at 20\,K; all atoms are at Wyckoff position $2a$.  Copper positions and $U_\text{iso}$ were fixed in the final step of the refinement.  The refined deuteration level is 92.495\,\%.}
  \begin{tabular}{lr@{.}lr@{.}lr@{.}lr@{.}l}\hline\hline
    Site & \multicolumn{2}{c}{$x$} & \multicolumn{2}{c}{$y$} & \multicolumn{2}{c}{$z$} & \multicolumn{2}{c}{$U_\text{iso}$}\\ \hline
    Cu1 & $-$0&0044 & $-$0&0183 & 0&99329 & 0&00135\\
    Cu2 & 0&50715 & 0&23160 & $-$0&0023 & 0&00135\\
    N & 0&2322(4) & 0&2391(9) & 0&4091(2) & 0&00135\\
    O1 & 0&8694(5) & 0&2356(14) & 0&8538(4) & 0&00135\\
    O2 & 0&3156(9) & $-$0&0137(13) & 0&8800(9) & 0&00135\\
    O3 & $-$0&3092(9) & $-$0&0234(13) & 0&1221(9) & 0&00135\\
    O4 & 0&2069(4) & 0&2354(14) & 0&2231(4) & 0&00135\\
    O5 & 0&3848(7) & 0&1227(7)&  0&4945(5) & 0&00135\\
    O6 & $-$0&1013(7) & $-$0&1340(8) & $-$0&4951(5) & 0&00135\\
    D1 & 0&8958(5) & 0&2381(13) & 0&7206(4) & 0&00135\\
    D2 & 0&3010(8) & 0&0005(12) & 0&7396(8) & 0&00135\\
    D3 & $-$0&2850(8) & $-$0&0347(12) & 0&2618(7) & 0&00135\\ \hline\hline
  \end{tabular}
\end{table}

\begin{table}[tb]
  \caption{\label{NPDHRPT1}Refined atomic positions in \rouaD\ from neutron powder diffraction on HRPT combining 1.886- and 2.449-\AA\ data collected at 10\,K; all atoms are at Wyckoff position $2a$.  The refined Cu positions were taken from a refinement of 20-K Wombat data which is not summarized here, $U_\text{iso}$ is taken from the Echidna refinement in Table~\ref{NPDEch1}, and the deuteration level of 92.495\,\%\ from the Echidna refinement was used.}
  \begin{tabular}{lr@{.}lr@{.}lr@{.}lr@{.}l}\hline\hline
    Site & \multicolumn{2}{c}{$x$} & \multicolumn{2}{c}{$y$} & \multicolumn{2}{c}{$z$} & \multicolumn{2}{c}{$U_\text{iso}$}\\ \hline
    Cu1 & $-$0&0030 & 0&0 & 0&9890 & 0&00135\\
    Cu2 & 0&50860 & 0&248 & $-$0&0027 & 0&00135 \\
    N & 0&2320(7) & 0&2545(18) & 0&4096(6) & 0&00135 \\
    O1 & 0&8719(10) & 0&251 & 0&8578(10) & 0&00135 \\
    O2 & 0&3087(18) & 0&004(3) & 0&8831(19) & 0&00135 \\
    O3 & $-$0&3075(20) & $-$0&007(3) & 0&1250(19) & 0&00135 \\
    O4 & 0&2050(9) & 0&251(4) & 0&2250(8) & 0&00135 \\
    O5 & 0&3815(14) & 0&1389(14) & 0&4946(12) & 0&00135 \\
    O6 & $-$0&1036(13) & $-$0&1162(16) & $-$0&5006(12) & 0&00135 \\
    D1 & 0&8940(10) & 0&256(3) & 0&7239(9) & 0&00135 \\
    D2 & 0&3030(18) & 0&019(3) & 0&7384(16) & 0&00135 \\
    D3 & $-$0&2849(17) & $-$0&015(3) & 0&2612(16) & 0&00135 \\ \hline\hline
  \end{tabular}
\end{table}

\begin{table}[tb]
  \caption{\label{NXDKoalaAtom}Refined atomic positions in \rouaH\ from neutron Laue diffraction on Koala2 at 10\,K; all atoms are at Wyckoff position $2a$.}
  \begin{tabular}{lr@{.}lr@{.}lr@{.}lr@{.}l}\hline\hline
    Site & \multicolumn{2}{c}{$x$} & \multicolumn{2}{c}{$y$} & \multicolumn{2}{c}{$z$} & \multicolumn{2}{c}{$U_\text{iso}$}\\ \hline
    Cu1 & $-$0&0016(6) & 0&0006(4) & 0&9929(3) & 0&0020(2)\\
    Cu2 & 0&5095(4) & 0&2520(4) & $-$0&0013(2) & 0&0021(2) \\
    N & 0&2338(3) & 0&2611(3) & 0&4084(2) & 0&0038(2) \\
    O1 & 0&8664(5) & 0&2545(5) & 0&8553(3) & 0&0039(3) \\
    O2 & 0&3126(7) & 0&0073(4) & 0&8767(4) & 0&0040(4) \\
    O3 & $-$0&3101(7) & $-$0&0043(4) & 0&1205(4) & 0&0031(4) \\
    O4 & 0&2071(5) & 0&2564(5) & 0&2245(3) & 0&0044(3) \\
    O5 & 0&3868(7) & 0&1407(4) & 0&4948(4) & 0&0073(4) \\
    O6 & $-$0&1013(7) & $-$0&1149(4) & $-$0&4963(4) & 0&0068(4) \\
    H1 & 0&8940(12) & 0&2591(13) & 0&7217(8) & 0&0208(8) \\
    H2 & 0&3018(16) & 0&0224(9) & 0&7358(11) & 0&0176(11) \\
    H3 & $-$0&2820(14) & $-$0&0158(10) & 0&2601(9) & 0&0162(10) \\ \hline\hline
  \end{tabular}
\end{table}

\section{Supplemental Material\label{supp}}

As ancillary files to this arXiv submission, we provide the following describing our crystal and magnetic structure refinements:

\noindent\begin{tabular}{l}
\verb+Echidna_20K_2p44A_rouaite.cif+\\
\verb+Echidna_3p2K_2p44A_rouaite.mcif+\\
\verb+HRPT_10K_1p89A_rouaite.cif+\\
\verb+HRPT_10K_2p45A_rouaite.cif+\\
\verb+Koala2_10K_Laue_rouaite.cif+
\end{tabular}

Each filename describes the diffractometer, measurement temperature in K, wavelength in \AA, and material.  The \verb+.mcif+ file describes the magnetic structure at 3.2\,K.

\bibliography{Rouaite}

\end{document}